\documentclass[a4paper,10pt]{article}
\pdfoutput=1 % if your are submitting a pdflatex (i.e. if you have
\usepackage{jheppub} % for details on the use of the package, please
                     \usepackage{mathpazo} % Use the Palatino font by default
\usepackage{amsmath}
\usepackage{amssymb}
\usepackage{hyperref}
\usepackage{float}
\usepackage{url}
\usepackage{mathtools}
\usepackage{caption}
\usepackage{subfig}
\usepackage[T1]{fontenc} % if needed

\title{\boldmath Development and characterization of six-gap glass MRPCs and feasibility study of a PET device}

\author[a,b,1]{M. Nizam,\note{Corresponding author.}}
\author[a,2]{B. Satyanarayana,\note{Corresponding author.}}
\author[a,3]{R. R. Shinde,}
\author[a,4]{G. Majumder}

\affiliation[a]{Tata Institute of Fundamental Research,\\Mumbai 400005, India}
\affiliation[b]{Homi Bhabha National Institute,\\Mumbai 400094, India}

\emailAdd{mohammad.nizam@tifr.res.in}
\emailAdd{bsn@tifr.res.in}
\emailAdd{rrs@tifr.res.in}
\emailAdd{gobinda@tifr.res.in}

\abstract{The Multigap Resistive Plate Chambers (MRPCs) provide
excellent timing as well as position resolutions at relatively low cost.
Therefore, they can be used in medical imaging applications such as
PET where precise timing is a crucial parameter of measurement. We
have designed and fabricated several six-gap glass MRPCs and extensively studied their performance. 
In this paper, we describe the fabrication and characterization of the detector,
the electronics and the data acquisition system of the setup. We present here
the result of our Time Of Flight (TOF) experiment using a radioactive source (${}^{22}Na$) hence to 
demonstrate their potential applications in medical imaging. We also present
the Geant4 based simulation results on the efficiency of our detector as a
function of the number of gaps and thickness of the converter material.}

\begin{document} 
%\linenumbers
\maketitle
\flushbottom

\section{Introduction}
\label{sec1}

The Multigap Resistive Plate Chamber (MRPC) is a modified version of RPC detector wherein 
the gas gap between the electrodes is further divided into multiple 
gaps by introducing electrically floating highly resistive plates. The MRPC
was first conceptualized and developed in 1996 \cite{mrpc1}. These detectors consist of
many highly resistive plates (e.g. glass) and very thin gas gap between them. The
high voltage is applied only on the outermost electrodes and the inner electrodes
are all electrically floating.The signals are readout from X- and Y-planes pickup
panels placed at the outside of anode and cathode electrodes. The time resolution of these detectors
 improves with narrower gas gap. Studies done by several groups have
shown a time resolution lower than 100 ps for various MRPC configurations \cite{mrpc2}.

\begin{figure}[H]%
    \centering
    \subfloat[]{{\includegraphics[width=7.5cm,height=7.5cm]{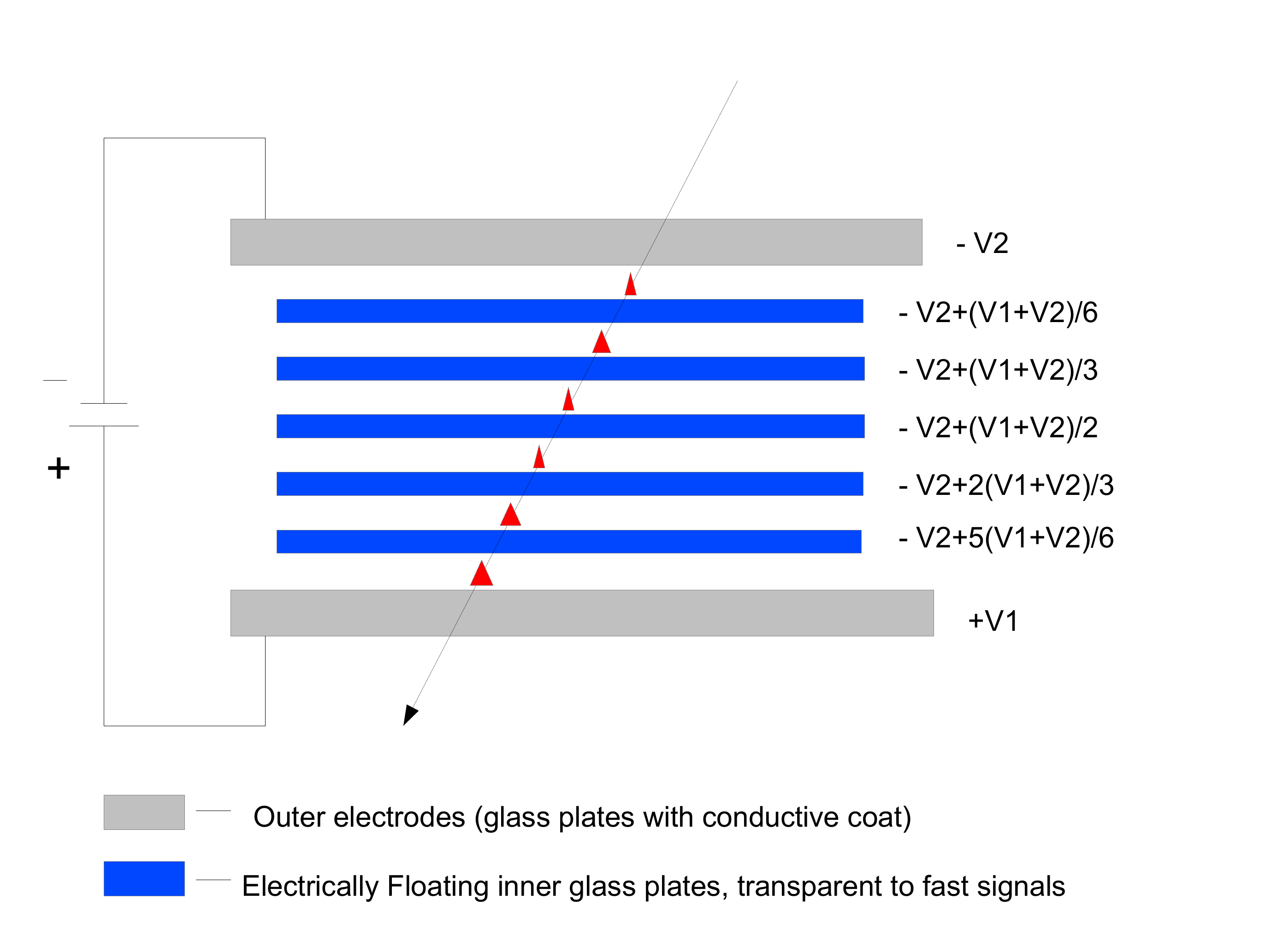} }}%
   % \qquad
    \subfloat[]{{\includegraphics[width=7.5cm,height=6.5cm]{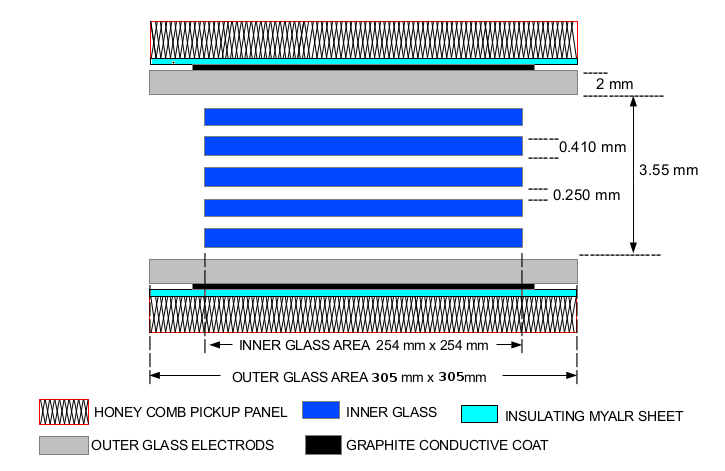} }}%
    \caption{(a) Potentials across the sub-gaps of an ideal MRPC detector and (b) a schematic diagram of our detector.}%
    \label{fig:mrpcdim}%
\end{figure}

The working principle of an MRPC is similar to that of a single gap RPC except that the electrically floating electrodes limits the avalanch to grow beyond
a certain level and provides less jitter in timing signal. Figure \ref{fig:mrpcdim}(a) illustrates an MRPC in the steady state condition. Excessive growth of 
the avalanche is limited by intermediate plates, and hence a higher electric field can
be applied to the detector operated in the avalanche mode, compared to that of a
single gap structure. This is advantageous in terms of the time resolution and rate
capability of the device.

The electrically floating interior resistive plates are maintained
at equal voltages due to the flow of positive ions and electrons between them.
The voltage across each gap is the same and on an average each gap
produces equal number of avalanches. The average net current to any of the internal
plates would be zero since there is an identical flow of electrons and ions in
each gas gap. The outer electrodes with graphite coat are transparent to the
fast signal generated by the avalanches inside each gas gap. The avalanche in
any of the gaps will induce a signal to the outermost electrodes, as the inner
electrodes are transparent to the fast signals. The fast signal is produced by
the flow of electrons towards the anode. The resultant signal is the summation
from all the gaps which is readout from  pickup panels placed 
at the outside of anode and cathode electrodes. MRPCs may consist of a
single stack or two stacks packed together \cite{mrpc3}. A higher operating voltage has to be applied due to larger over all gap compared to a single gap
RPC. MRPCs can have potential application in Positron Emission Tomography (PET) due to their precision time measurement \cite{K. Doroud,Sharifi,Belli} and good
spatial resolution \cite{Lshi}. Several glass MRPC detectors have
been developed to find potential application in TOF detectors, medical imaging
etc. Here we present the development and fabrication procedure of the optimized
design, characterization and the experimental setup of TOF-PET experiment, the trigger and data acquisition system, data analysis and Geant4 simulations of efficiency.

\section{Fabrication and Characterization of MRPCs} \label{fab}
\subsection{Fabrication}

We have constructed several six-gap glass MRPCs of dimensions 305 mm $\times$
305 mm $\times$ 7.5 mm. A schematic of the detector geometry with dimensions of
various components is shown in figure \ref{fig:mrpcdim}(b). The dimensions of internal 
glass plates are 256 mm $\times$ 256 mm $\times$ 0.410 mm. Glass sheets 
of 2 mm thickness, coated with a conductive layer of graphite paint of NEROLAC brand,
were used for the outer electrodes. The surface resistivity of the conductive coat was in the range of 0.5 - 1 M$\Omega$/$\square$. Two sided non conducting adhesive tapes 
were used on both sides of a mylar sheet to make small circular spacers of diameter
4 mm and thickness $\approx$ 250 $\mu$m. Twenty five spacers were used to maintain gas gaps.

\begin{figure}[H]%
    \centering
    \subfloat[]{{\includegraphics[width=6cm,height=3.5cm]{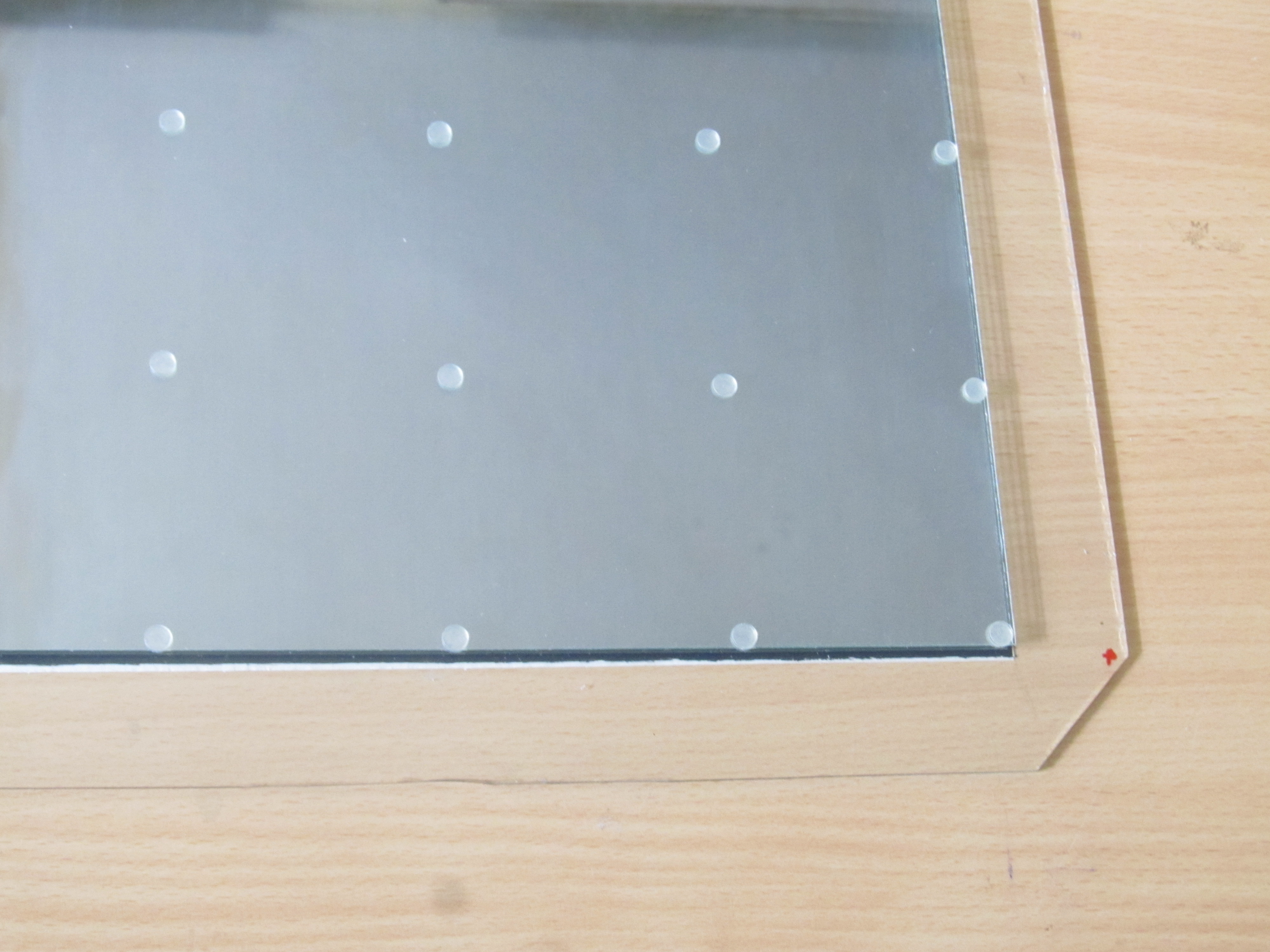} }}%
   % \qquad
    \subfloat[]{{\includegraphics[width=6cm,height=3.5cm]{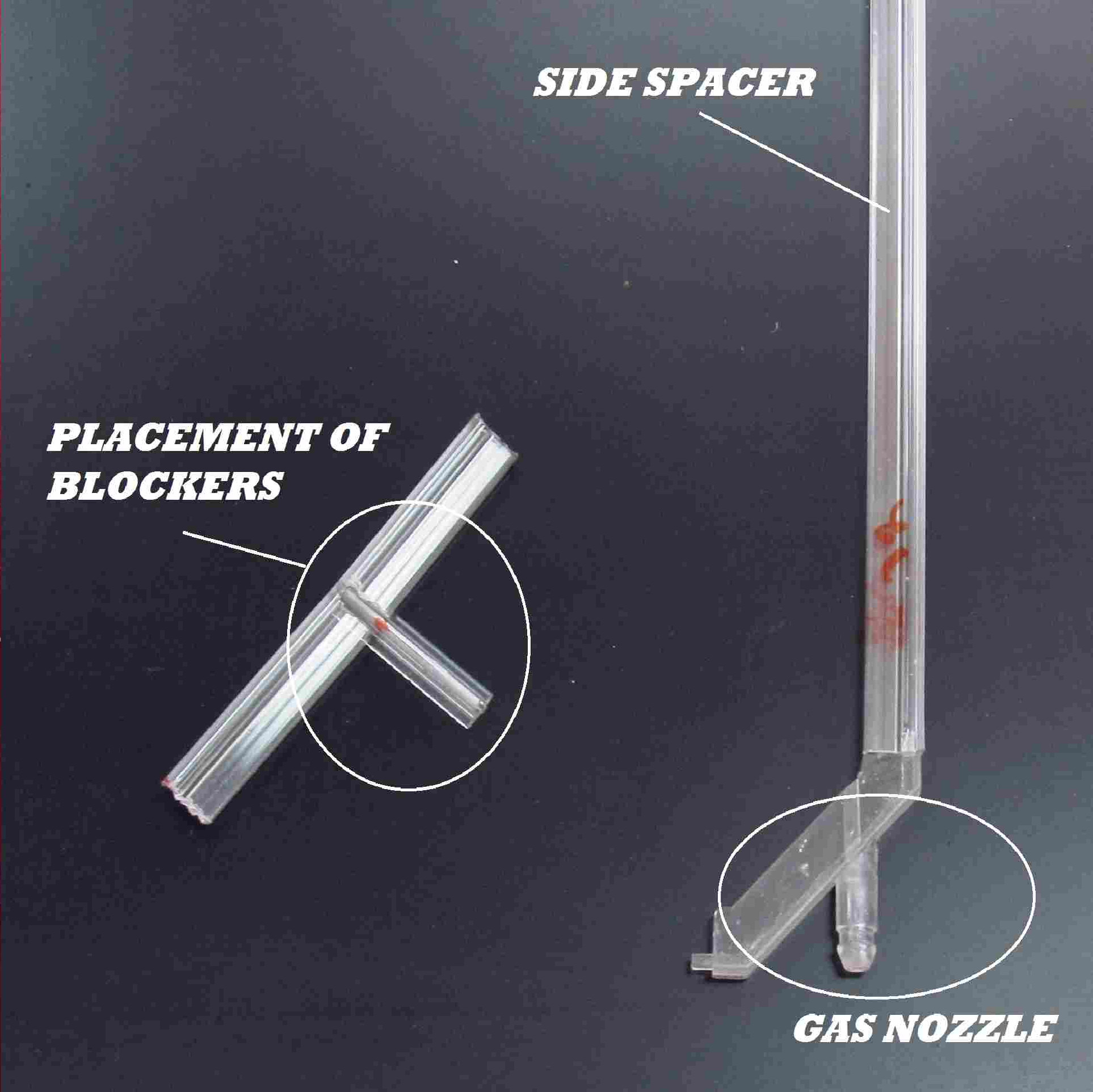} }} \\
     
    \subfloat[]{{\includegraphics[width=6cm,height=3.5cm]{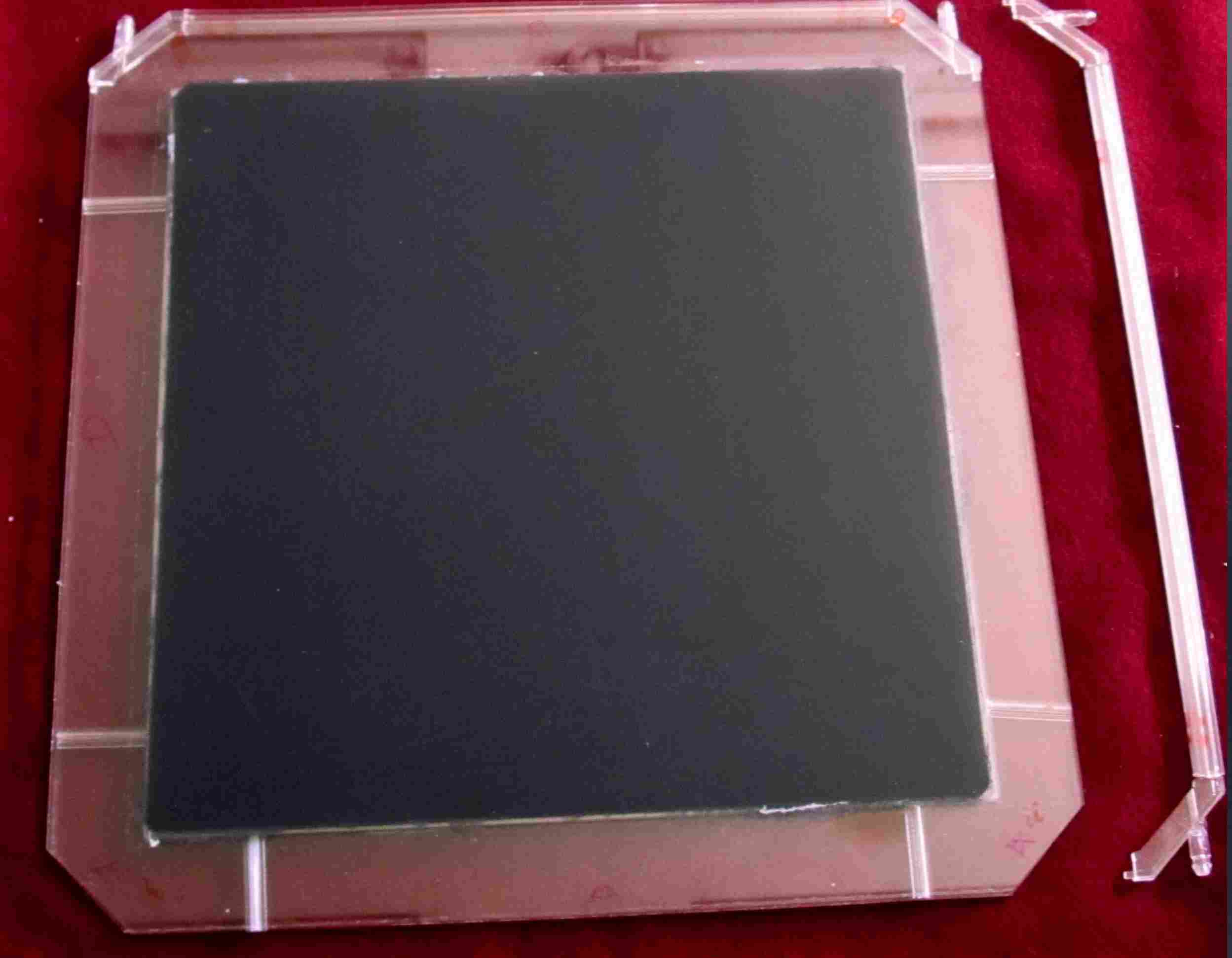} }}%
    %\qquad
    \subfloat[]{{\includegraphics[width=6cm,height=3.5cm]{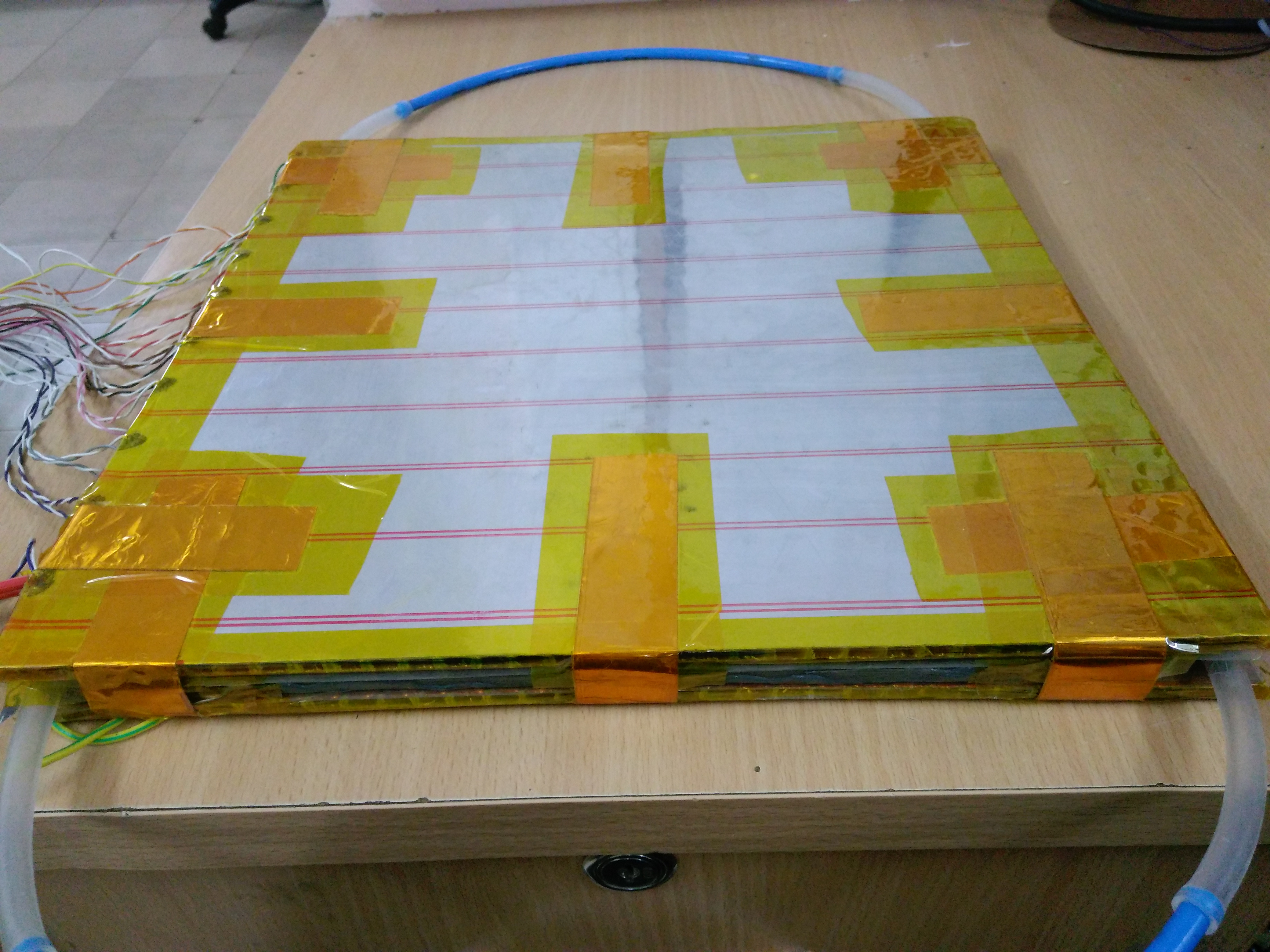} }}%
    \caption{(a) Placement of spacers, (b) blockers and side spacers, (c) placement of blockers and side spacers and (d) a fully assembled
    MRPC with pickup panels.}%
    \label{fig:fab}%
\end{figure}

Figure \ref{fig:fab}(a) shows placement of the spacers. There is a gap of around 2.7 cm
between the edges of external and internal electrodes. The gas mixture might end up flowing through 
that path of thickness 3.55 mm, instead of flowing through a very narrow (0.250
mm) gas gap, which would offer much higher resistance to the gas flow. In
order to ensure a proper flow of gas mixture through gap between the internal plates, we introduced blockers
at appropriate places (one each near the gas inlets and two each near the gas
outlets). Figures \ref{fig:fab}(b) and \ref{fig:fab}(c) shows the design and placement of the blockers, side spacers and gas nozzels respectively. 
The gas mixture has been optimized to $R134a$ (91.2 \%), $C_4 H_{10}$ (4.8\%) and $SF_6$
(4\%). The pickup panels consist of plastic honeycomb material laminated with copper strips of width 2.8 cm placed on both the sides of an MRPC orthogonal to each other. 
Figure \ref{fig:fab}(d) shows a fully assembled detector. 
Further details of our detector can be found in \cite{mrpc4}. 

\subsection{Characterization}
\begin{figure}[H]%
    \centering
    \subfloat[]{{\includegraphics[width=7.5cm,height=3.2cm]{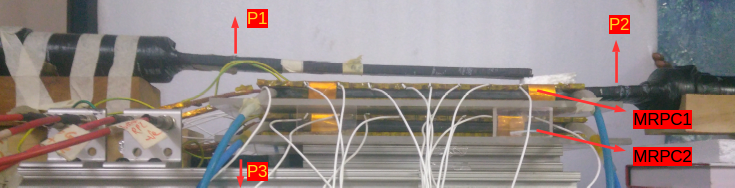} }}%
   % \qquad
    \subfloat[]{{\includegraphics[width=7.5cm,height=3.2cm]{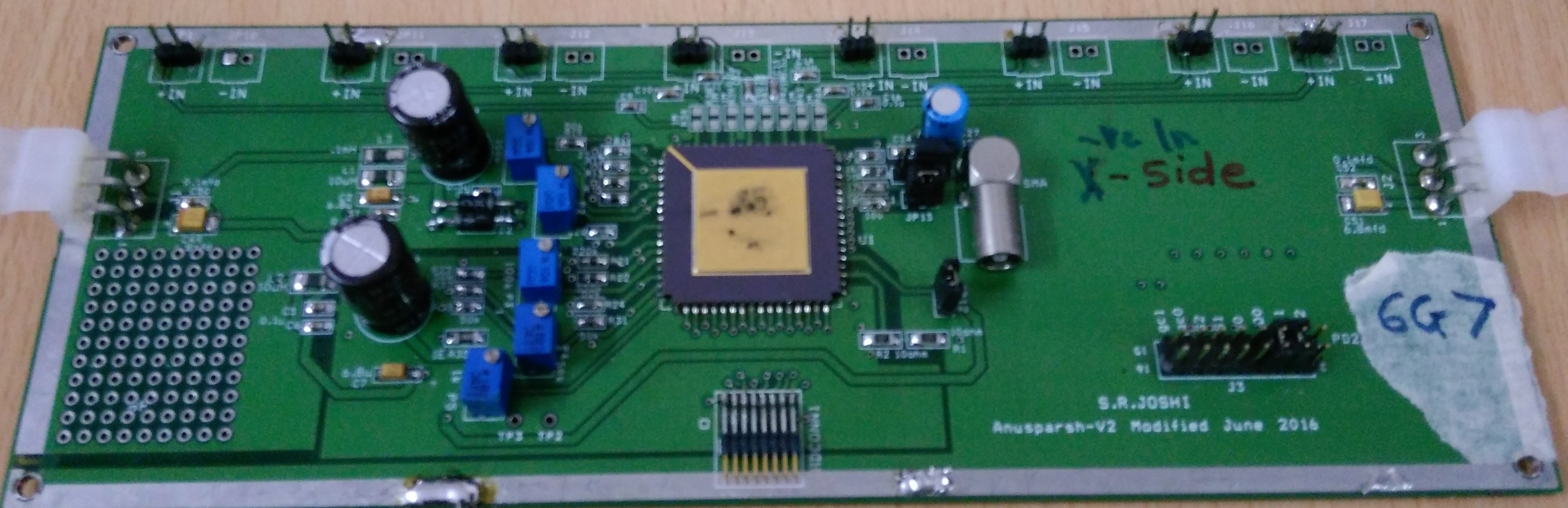} }}
    \caption{(a) A view of the Characterization setup and (b) the Anusparsh board.}%
    \label{fig:char1}%
\end{figure}

\begin{figure}[H]
\centering
\includegraphics[width=15cm,height=6.5cm]{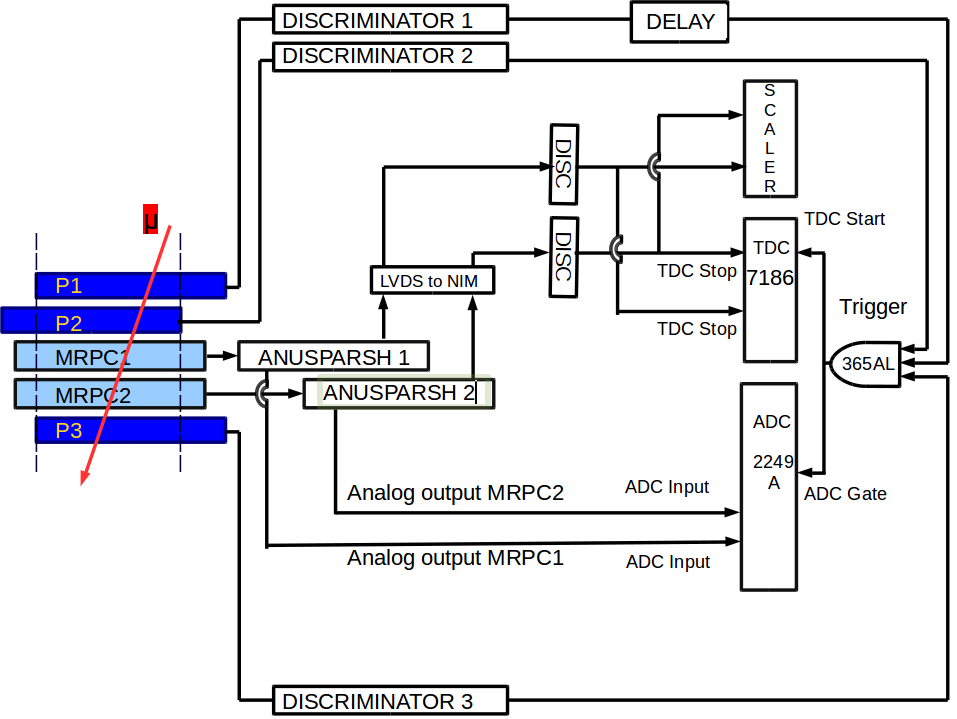}
\caption{DAQ of the characterization setup}
\label{fig:char2}
\end{figure}

We have characterized the two MRPCs used in our Time of Flight experiment. 
A cosmic ray muon telescope was setup using three scintillator paddles of 2 cm width. The two scintillator paddles were placed
above the two MRPCs under test and the third paddle was placed below the MRPCs. All the three paddles were 
aligned along the central strip of both MRPCs. The effective area of the telescope was  
25 cm $\times$ 2 cm. A view of the characterization setup is shown in figure \ref{fig:char1}(a). P1, P2 and P3 are the scintillator paddles
whose coincidence is given as the trigger to the Data Acquisition System (DAQ). A schematic of the DAQ system is shown in 
figure \ref{fig:char2}. The trigger is formed by the coincidence of discriminated signals of P1, P2 and P3.
The signals from MRPC1 and MRPC2 are recorded with this trigger. We need both digital as well as analog outputs from the preamplifiers 
for correcting time walk of Time to Digital Converter (TDC) data. Anusparsh boards \cite{anu} are used as preamplifiers to obtain time and charge 
information simultaneously. Anusparsh is a front end ASIC designed for the ICAL experiment. It is an 8-channel amplifier and 
discriminator. It also provides analog output of the amplifier stage for a selected channel. The amplifier comprises of a 
regulated cascode transcendence amplifier, followed by two stages of differential amplifier. The threshold is common to all 
discriminator channels. Outputs of the discriminators are Low Voltage Differential Signals (LVDS). The charge information is 
obtained from the analog output of the Anusparsh and used for the calibration of the TDC data. A picture of the Anusparsh boards 
designed and fabricated at TIFR is shown in figure \ref{fig:char1}(b).
The analog outputs are directly fed to the ADC inputs to get the charge information.
The LVDS outputs are first converted into NIM signals. These NIM signals are discrimated and fed into the TDC module to get timing
information of the MRPC signals with respect to the trigger and into the scaler module to get individual noise rates and efficiencies 
of MRPC1 and MRPC2.

\begin{figure}[H]%
    \centering
    \subfloat{{\includegraphics[width=5.2cm,height=6.1cm]{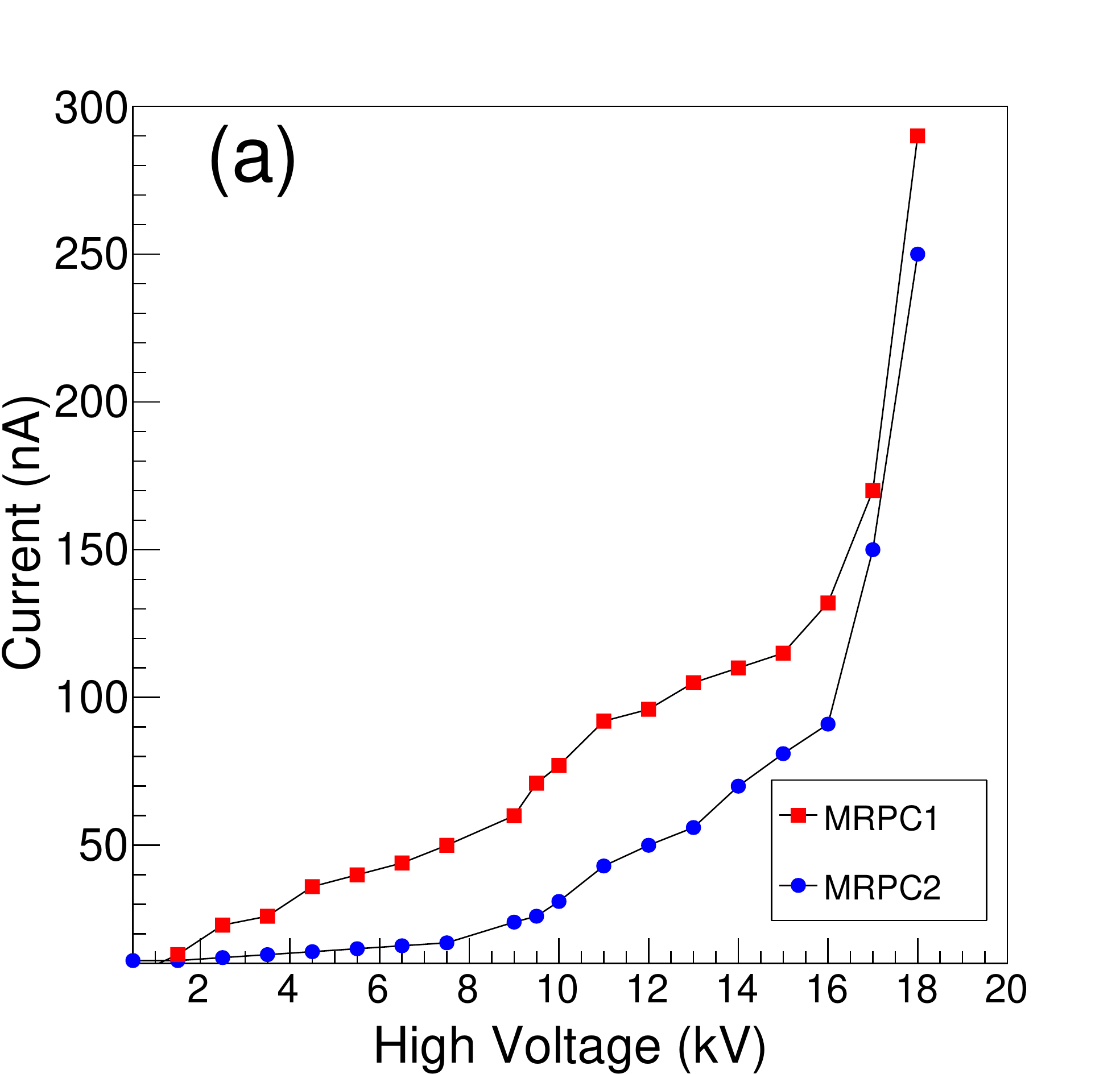} }}
    \subfloat{{\includegraphics[width=5.2cm,height=6.1cm]{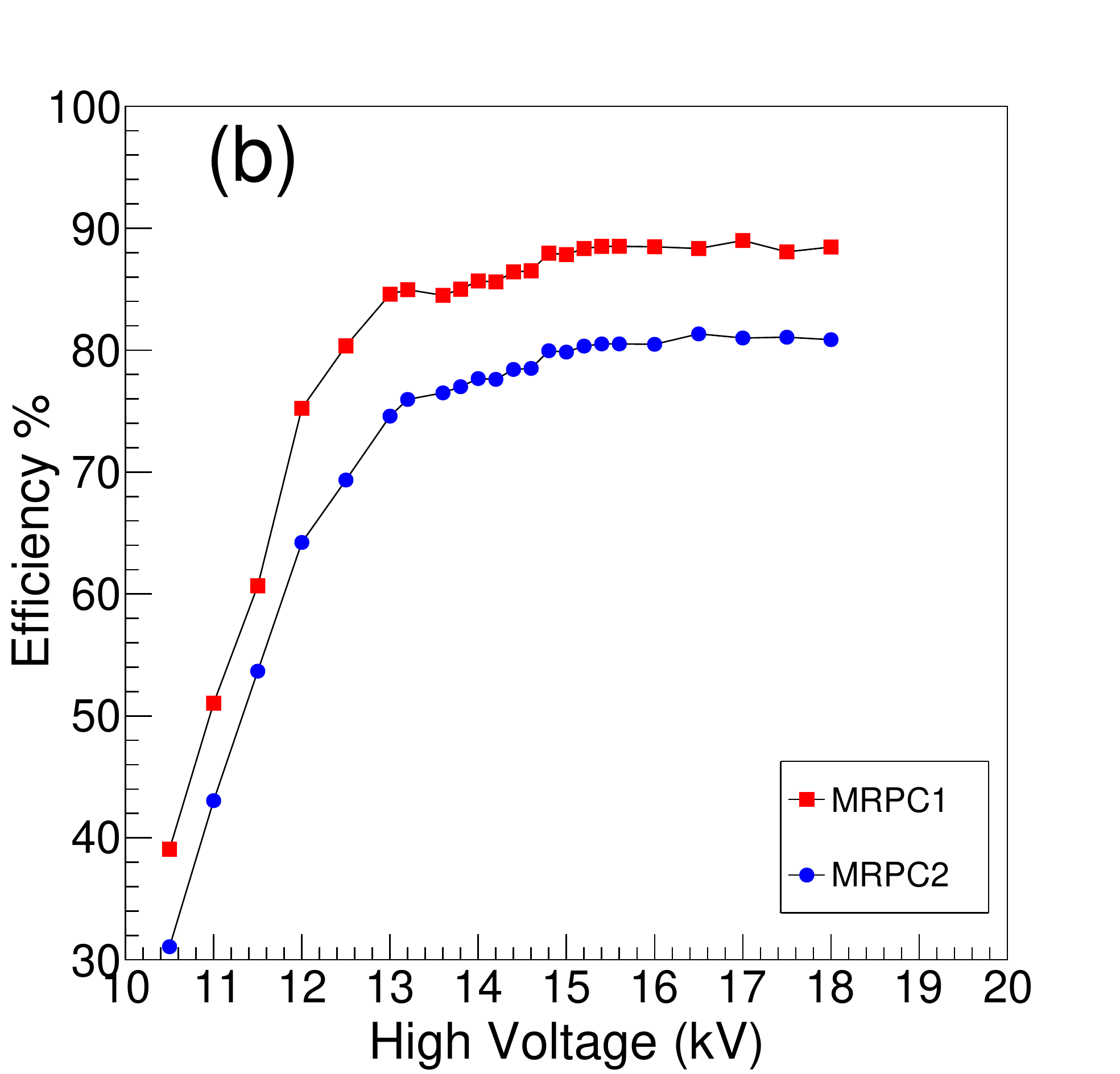} }}
    \subfloat{{\includegraphics[width=5.2cm,height=6.1cm]{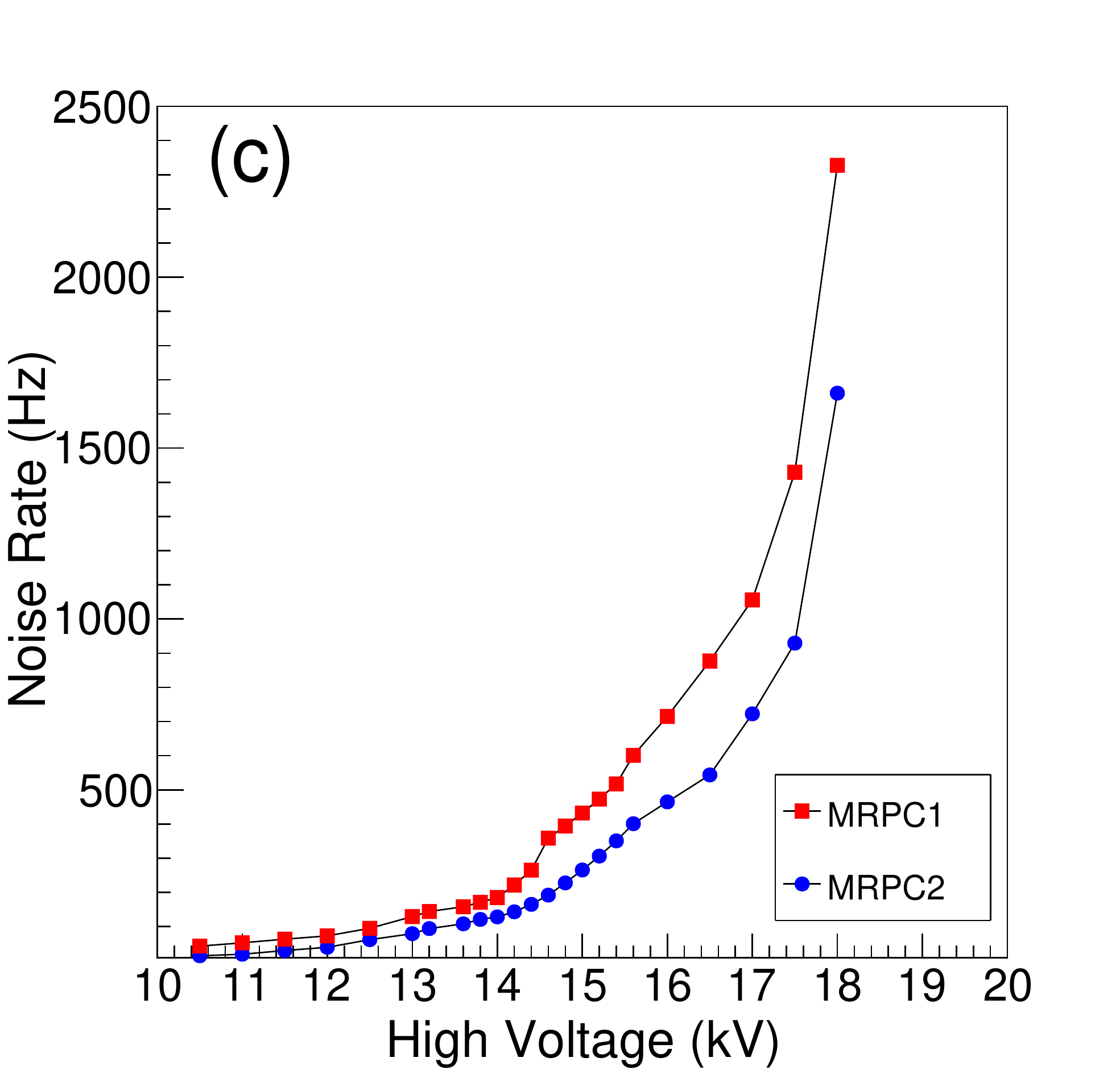} }} \\
    \subfloat{{\includegraphics[width=5.2cm,height=6.1cm]{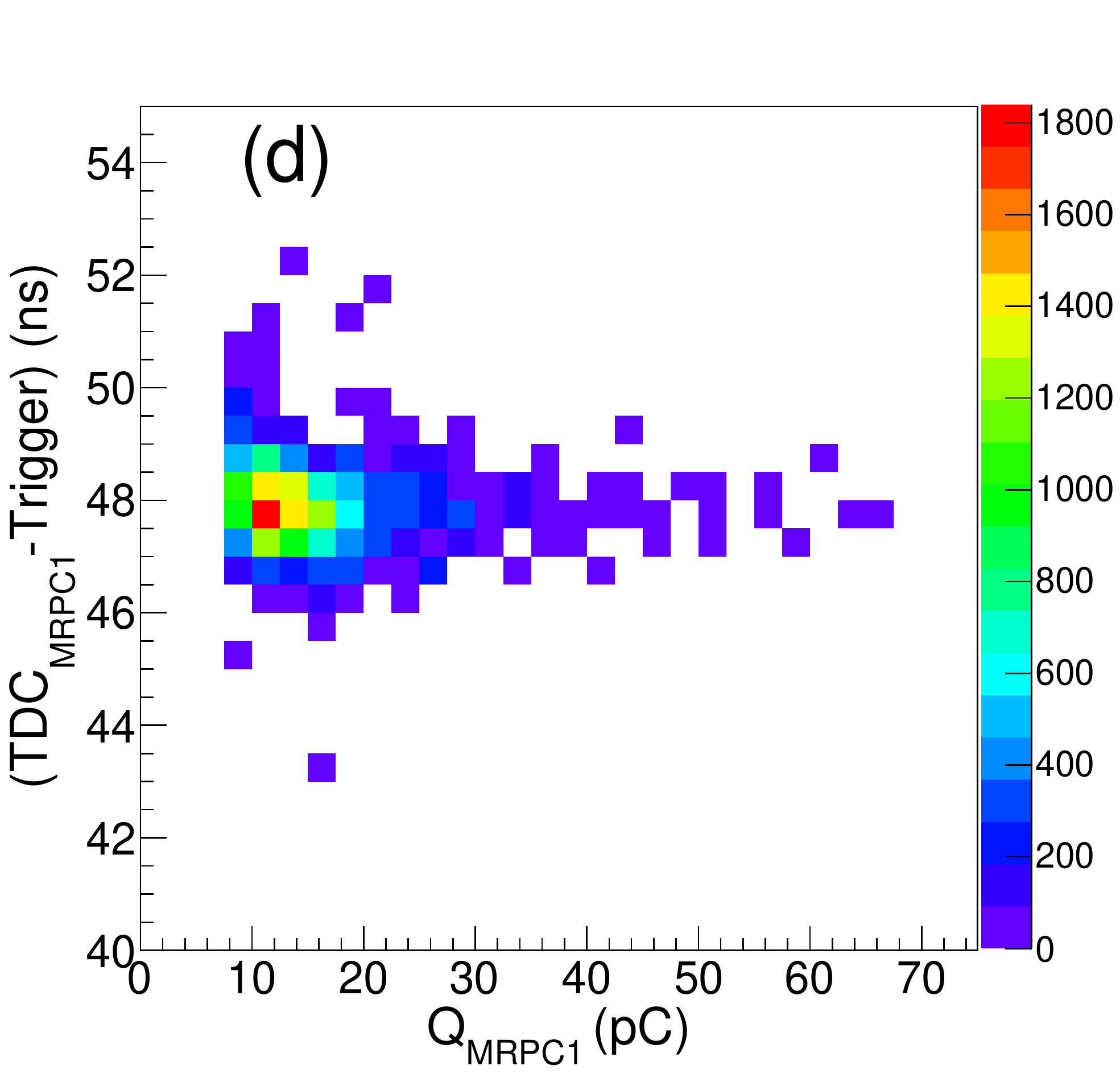} }}
    \subfloat{{\includegraphics[width=5.2cm,height=6.1cm]{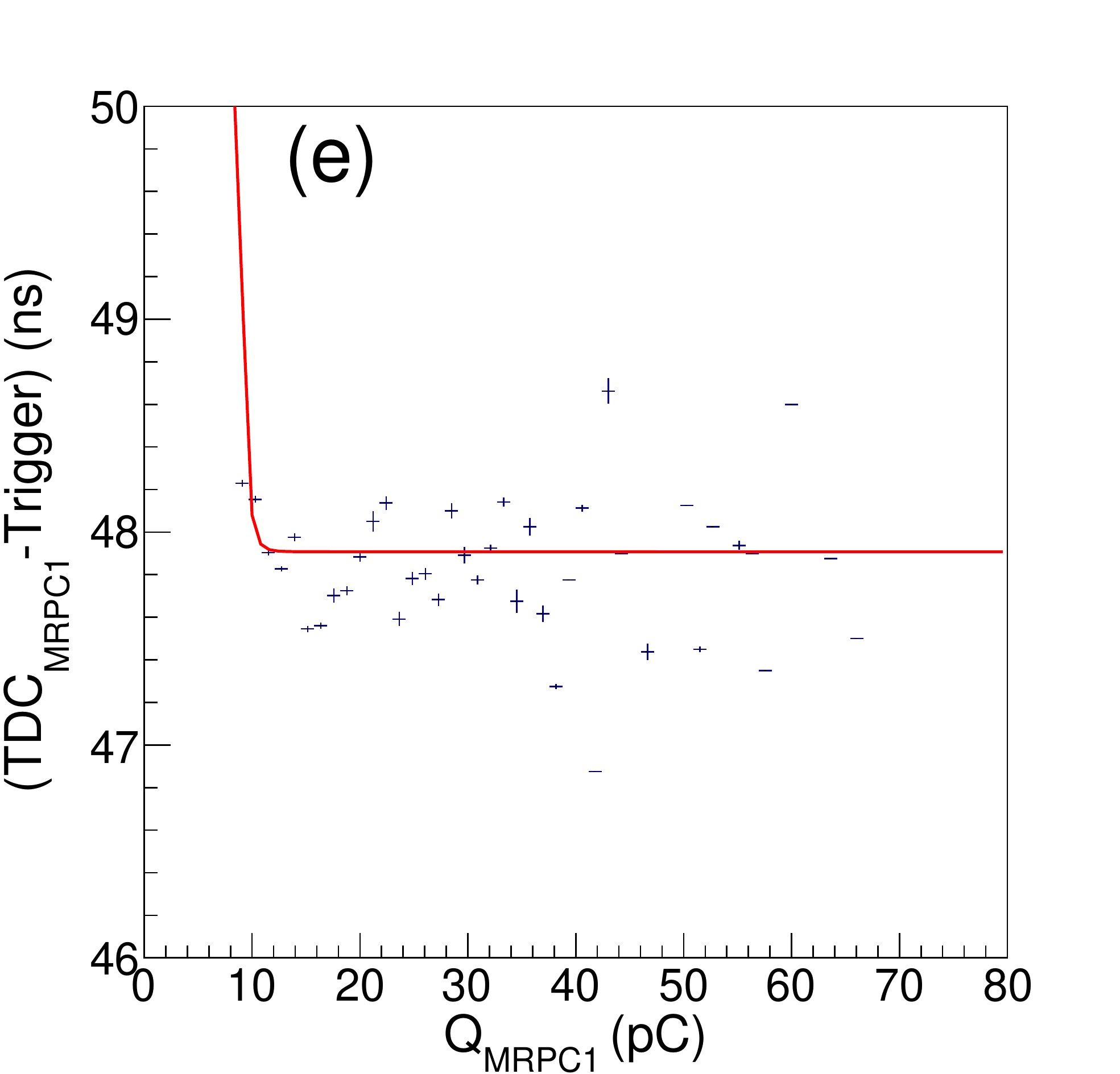} }} 
    \subfloat{{\includegraphics[width=5.2cm,height=6.1cm]{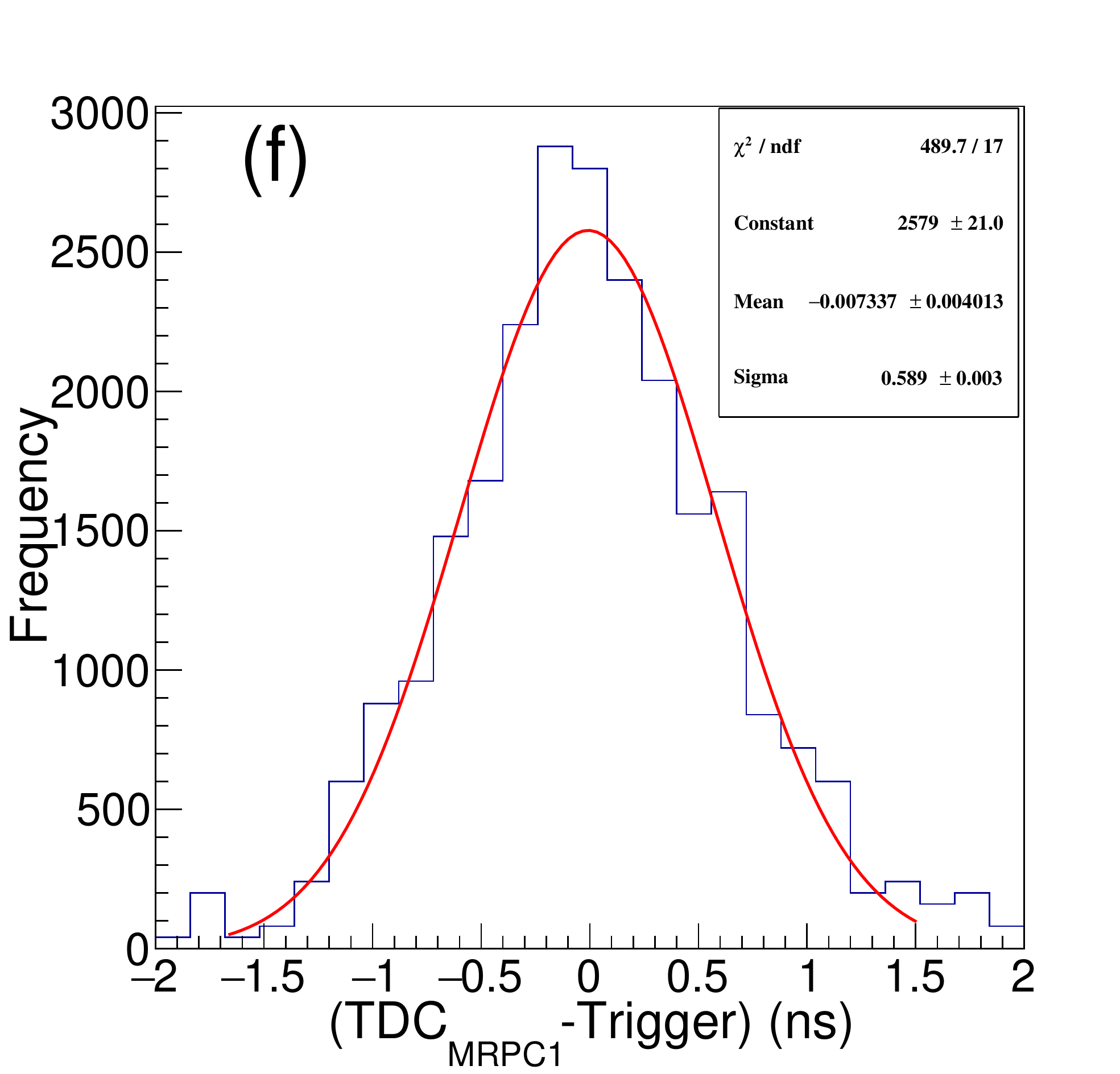} }} \\
    \subfloat{{\includegraphics[width=5.2cm,height=6.1cm]{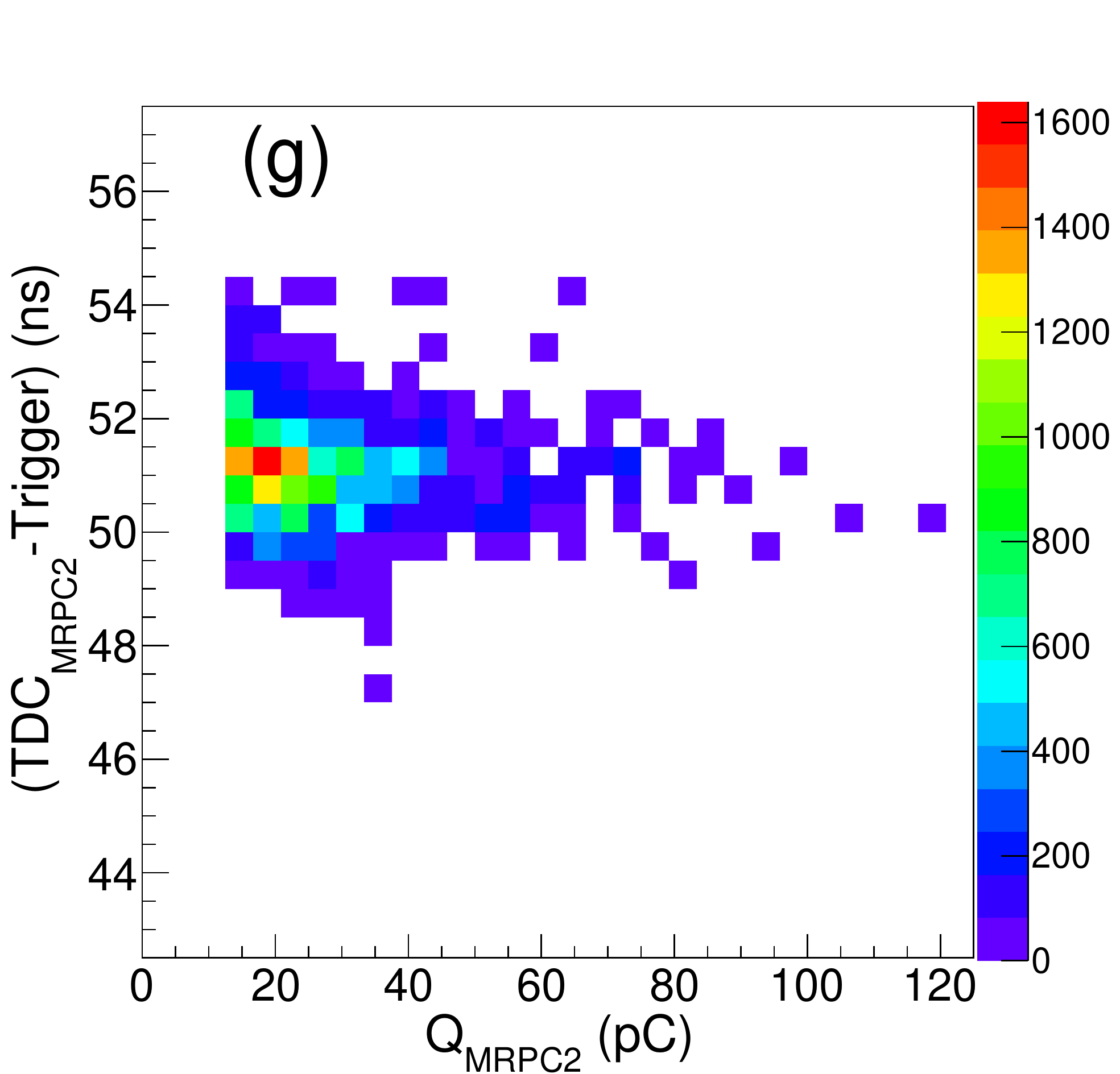} }}
    \subfloat{{\includegraphics[width=5.2cm,height=6.1cm]{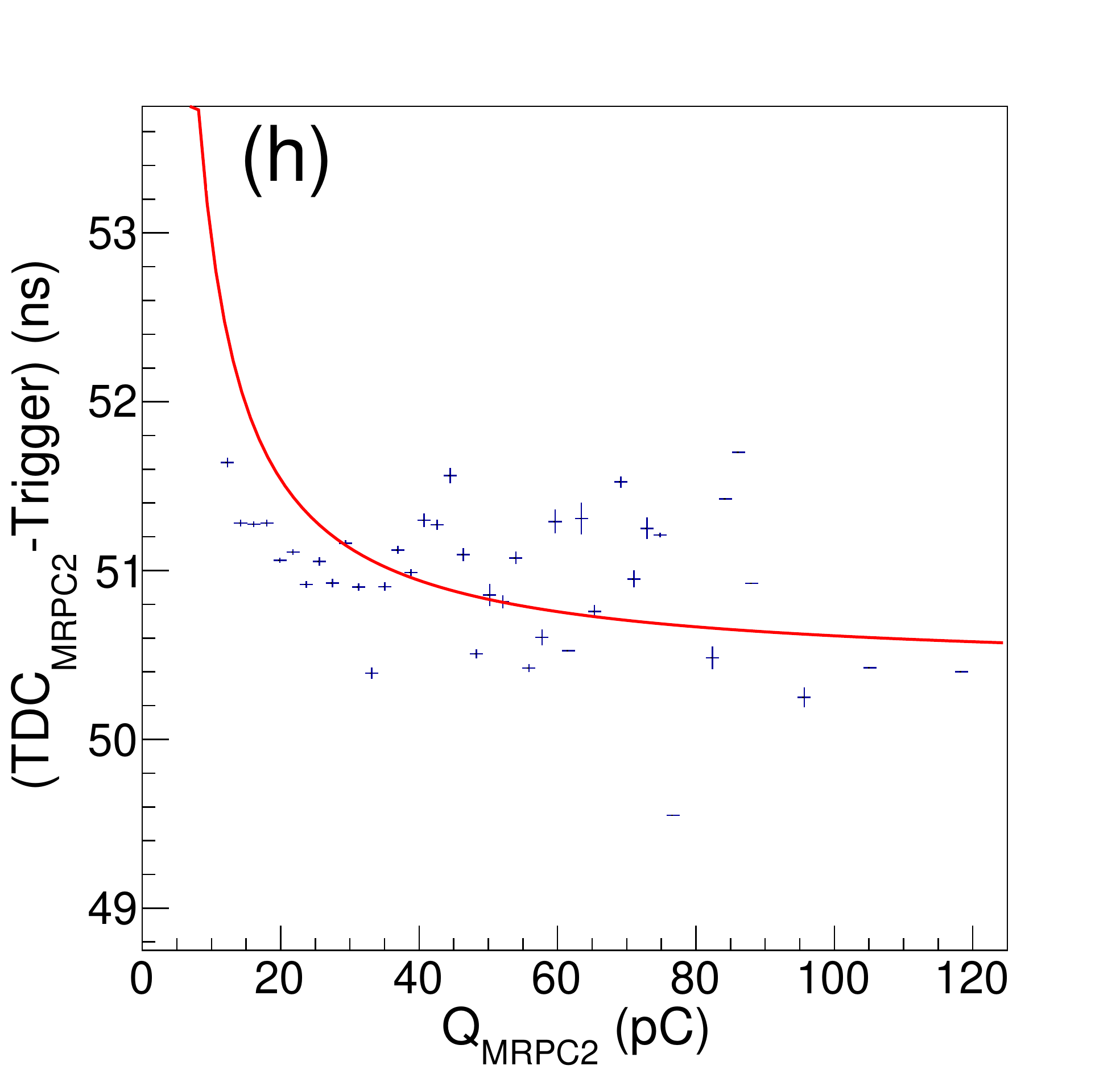} }} 
    \subfloat{{\includegraphics[width=5.2cm,height=6.1cm]{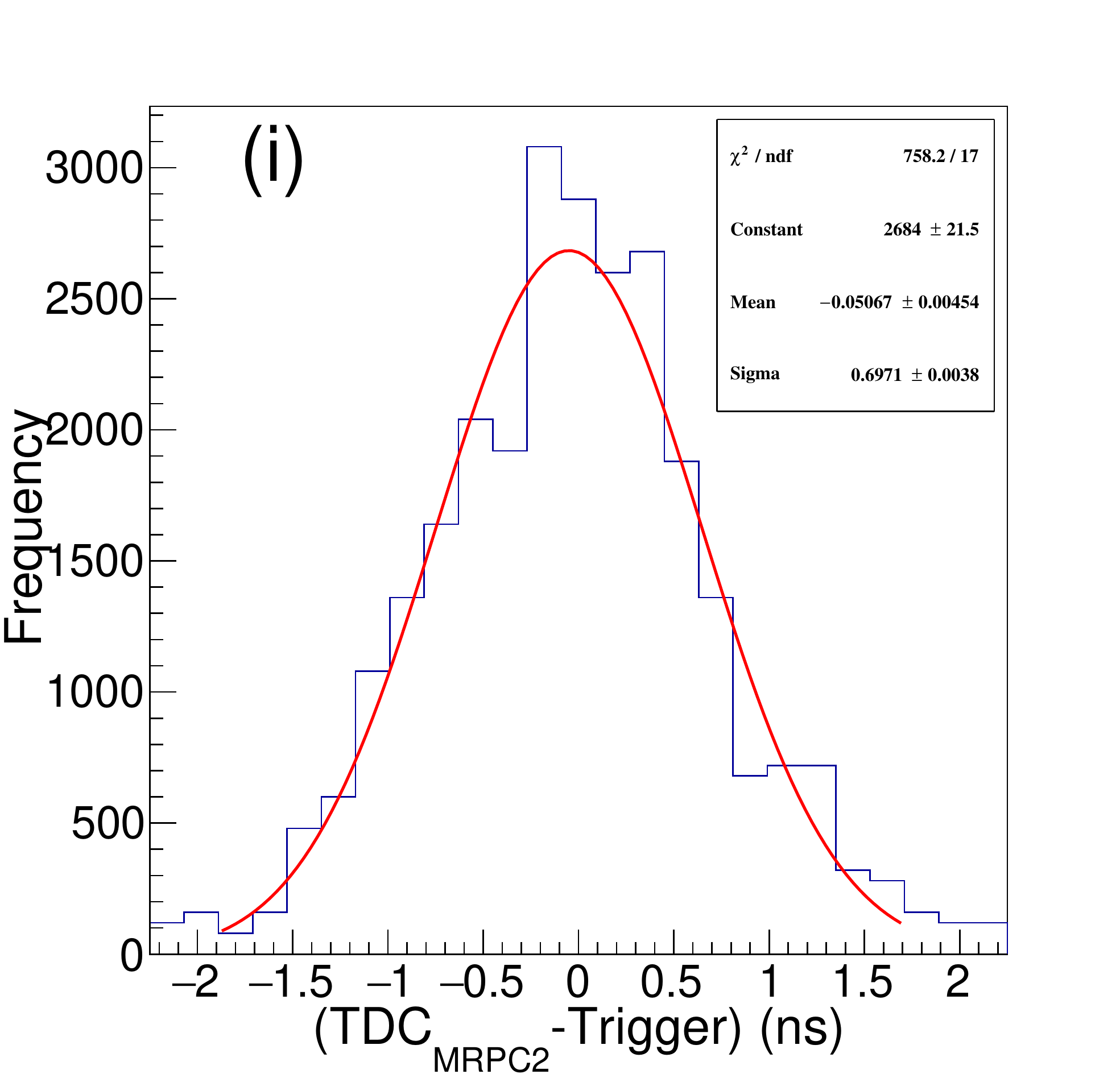} }}
    \caption{Characterization plots (a) Noise rates, (b) chamber current (I) vs applied high voltage (V), (c) efficiency versus applied high voltage, (d) TDC versus QDC histogram of MRPC1, (e) Profile histogram of MRPC1, (f) the time walk corrected TDC distribution of MRPC1, (g) TDC versus QDC histogram of MRPC2, (h) Profile histogram of MRPC2 and (i) the time walk corrected TDC distribution of MRPC2.}%
\label{fig:charplots}%
\end{figure}

The noise rates verus applied high voltages, I-V charcteristics and the muon detection efficiencies versus applied high voltages of MRPC1 and MRPC2
are shown in figures \ref{fig:charplots}(a,b and c) respectively. We also studied time resolutions of MRPC1 and MRPC2 with this 
characterization setup using the cosmic muons. The TDC data of individual MRPCs were corrected by using the charge information of 
each event obtained by Anusparsh boards. Figures \ref{fig:charplots}(d and g) and \ref{fig:charplots}(e and h) show the TDC counts versus QDC counts 
histograms and profile histogram of MRPC1 and MRPC2. The profile histograms were fitted to a function exp$[-p_0 /x + p_1] + p_2$. 
The TDC value of each event is then corrected 
by using the fit parameters $p_0 , p_1$ and $p_2$ and the QDC value. The calibration is done using the following equation
\begin{equation}
TDC_{corrected} = TDC_{raw} - T_{calib}
\label{eqn1}
\end{equation}
where $T_{calib}$ is the correction obtained using the fit parameters from the profile histogram. 
Figures \ref{fig:charplots}(f) and \ref{fig:charplots}(i) shows the corrected TDC distributions of MRPC1 and MRPC2 respectively. 
The corrected time
resolutions of MRPC1 and MRPC2 are $\sim$ 589 ps and $\sim$ 697 ps respectively which 
include $\sim$ 20 ps electronic jitter. Different characterization values of MRPC1 and MRPC2 are listed
in table \ref{table:chartab}.
\begin{table}[tbh]
\centering
\begin{tabular}{|c|c|c|c|c|}
\hline
  & Noise rate (Hz)& Current (nA) & Efficiency (\%)& $\sigma_{T}$ Corrected (ps)  \\
\hline
MRPC1 & $\sim$ 2350 & $\sim$ 290 & $\sim$ 88 & $\sim$ 589 \\ 
\hline
MRPC2 & $\sim$ 1700 & $\sim$ 250 & $\sim$ 80 & $\sim$ 697  \\
\hline
\end{tabular}
\caption{Noise rates, chamber currents, efficiencies and corrected timing resolutions of MRPC1 and MRPC2 at $\sim$ 18 kV operating 
voltage.}
\label{table:chartab}
\end{table}
\section{MRPC for Positron Emission Tomography (PET)}
We attempt to demonstrate the possible application of these detectors in medical imaging because of their excellent time resolution.
We have mounted two MRPCs horizontally and a radioactive source (${}^{22}Na$) is
placed asymmetrically between the two detectors. ${}^{22}Na$ emits a positron which
annihilates with an electron almost at rest and two gammas of 511 keV are
produced with opposite momenta. These photons are detected by the two detectors in coincidence with each other. Each MRPC has eight "X" strips
and eight "Y" strips of 2.8 cm width. Lines Of Response (LOR) can be obtained
by joining the hit coordinates. The time of flight information gives the exact
position of the source on the line of response. 
\subsection{The Experimental setup and DAQ}
\begin{figure}[H]%
    \centering
    \subfloat{{\includegraphics[width=11cm,height=8cm]{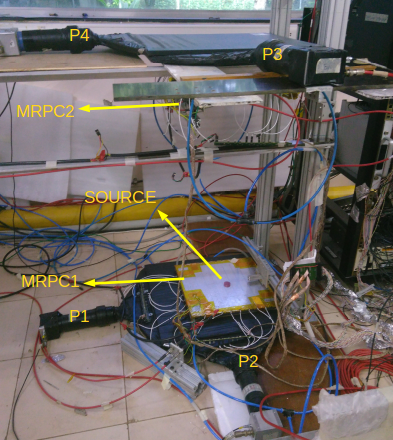} }}\\
    \subfloat{{\includegraphics[width=14cm,height=8cm]{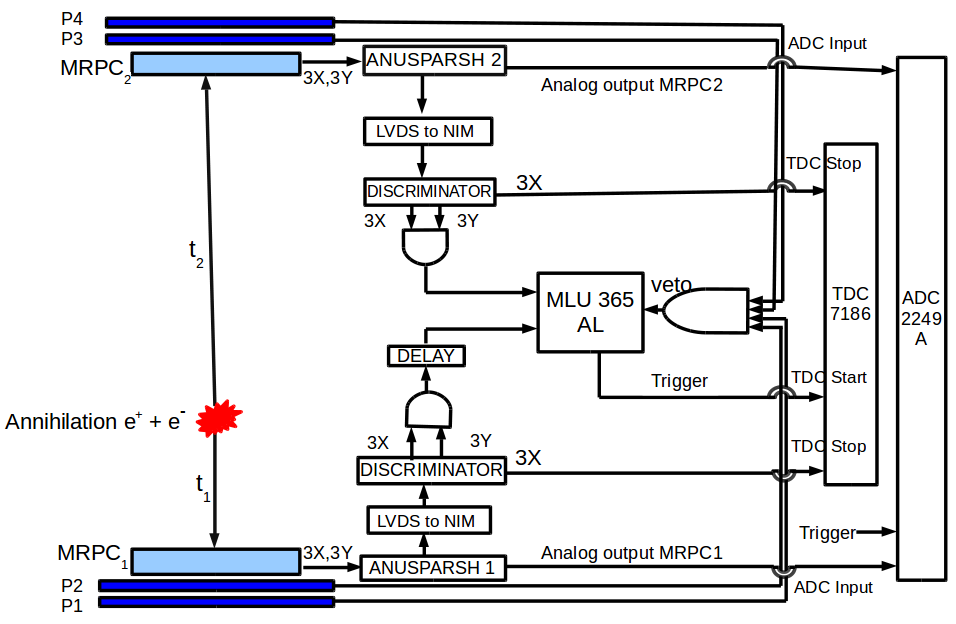} }}
    \caption{Top panel: A view of the setup. Bottom panel: Block diagram of the setup.}%
\label{fig:tofset}%
\end{figure}

A pair of scintillator paddles (P3 and P4) of dimensions 44 cm $\times$ 44 cm are 
placed above the top MRPC and another pair of scintillator paddles (P1 and P2) are placed below the bottom MRPC such that the MRPCs are well within the area of scintillator paddles. The coincidence signal of all four scintillator paddles is used as a veto to remove the cosmic muon background. The signals from pickup strips are readout by Anusparsh ASIC described in section 2. We read only three "X" and three "Y" central strips of each MRPC to form the trigger. The LVDS output of Anusparsh is converted to NIM signal. These are discriminated and fanned out for the MLU (LeCroy 365AL Dual 4-Fold Majority Logic Unit), Scalers and the TDC (Phillips Scientific 7186 Time-to-Digital Converter). In the coincidence unit path, the "X" strips and "Y" strips are ORed separately and the ORed
signals of X and Y planes are ANDed. The resultant AND output of each MRPC
are finally ANDed to form the trigger. In the TDC path, the discriminated NIM output of each strip is delayed by adding appropriate lenghts of cables such that the signal is well within the trigger window and then they are used as TDC stop. The analog outputs of Anusparsh boards are directly fed into the ADC inputs.  Figure \ref{fig:tofset} (bottom panel) illustrate the DAQ system. We have used only the "X" plane timing data of MRPCs to calculate the time of flight in this experiment. Figure \ref{fig:tofset} (top panel) shows a view of the experimental setup.
\subsection{Time of flight measurement}
\begin{figure}[H]%
    \centering
    \subfloat{{\includegraphics[width=7.7cm,height=5.3cm]{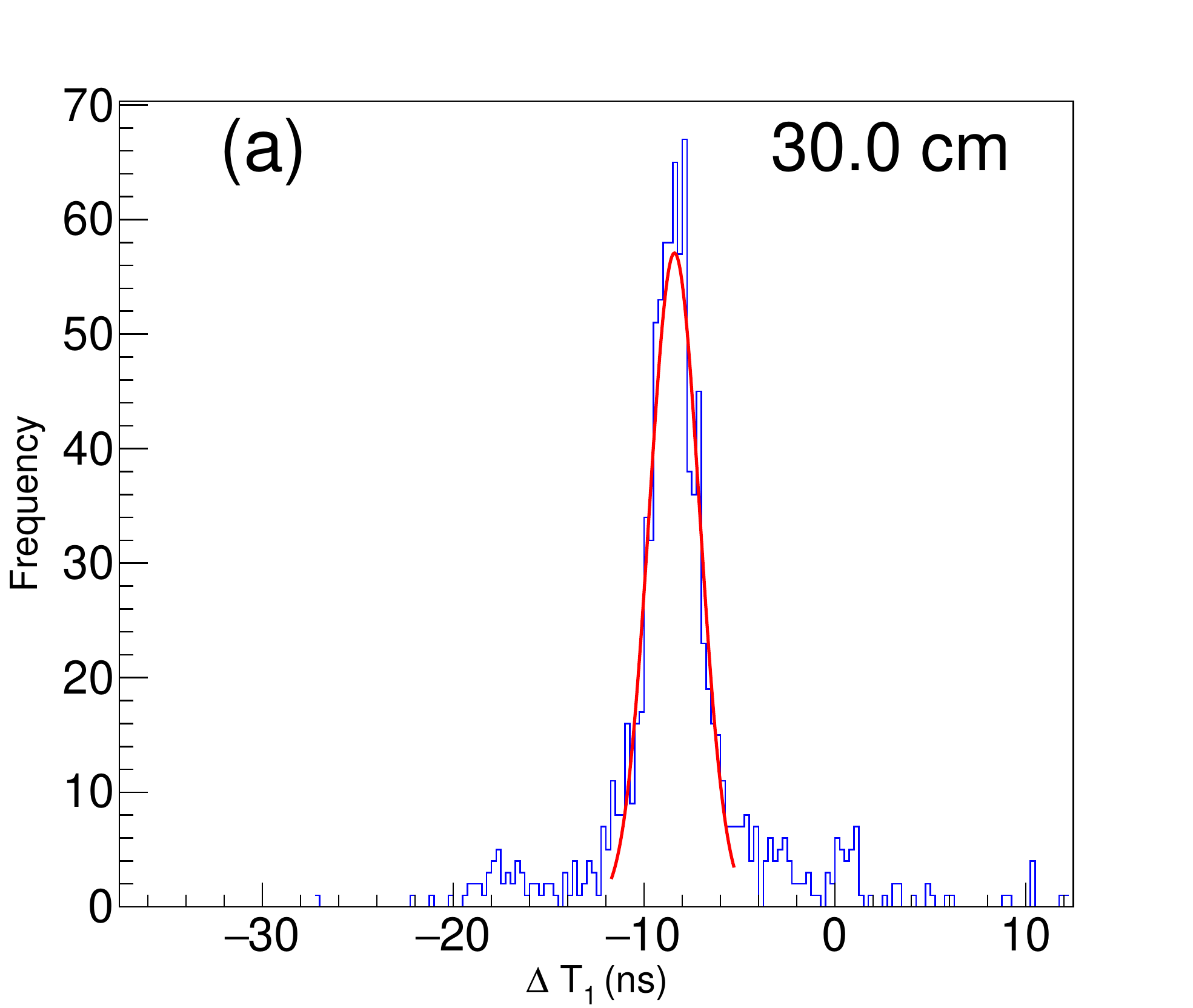} }}
    \subfloat{{\includegraphics[width=7.7cm,height=5.3cm]{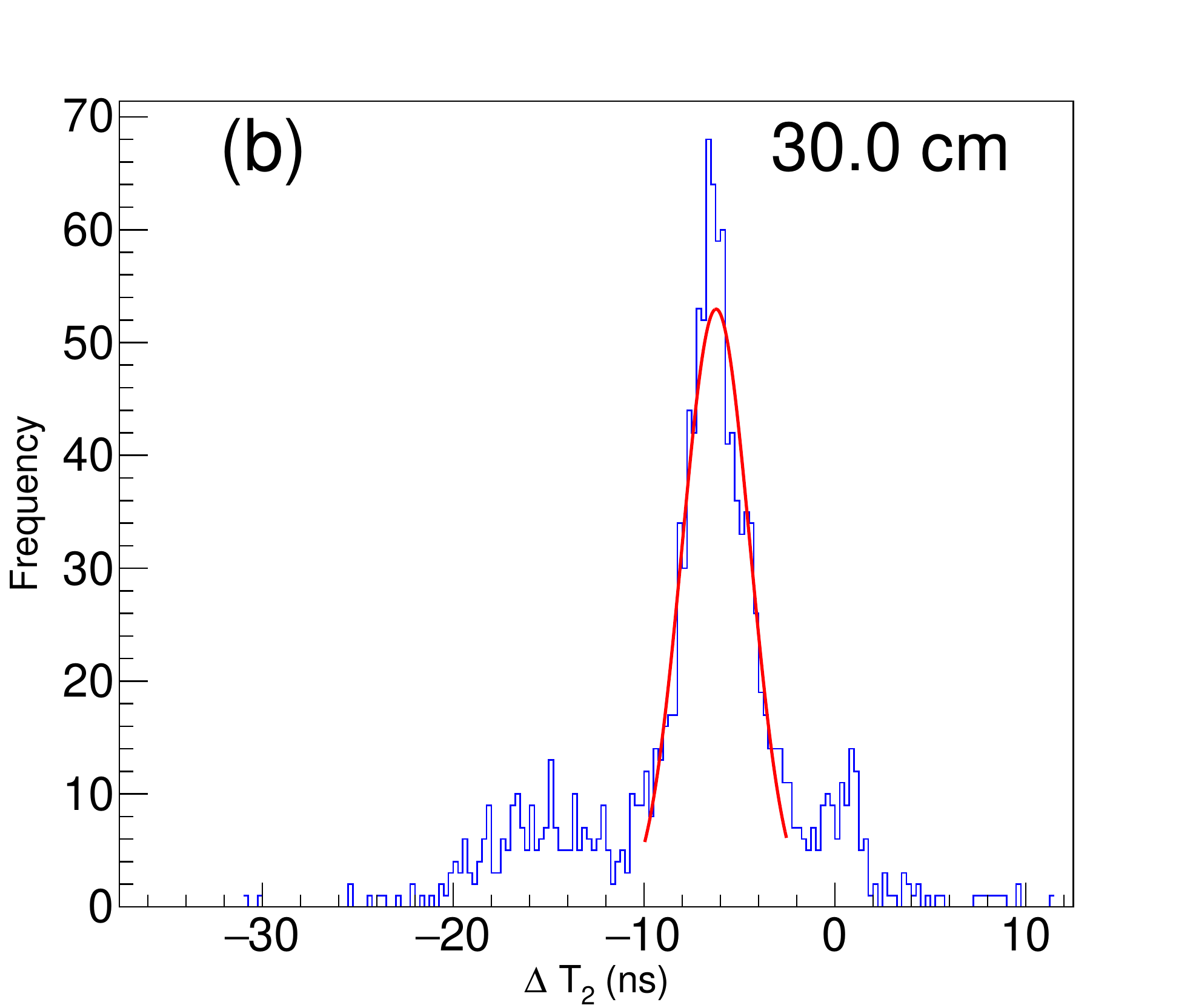} }}\\
    \subfloat{{\includegraphics[width=7.7cm,height=5.3cm]{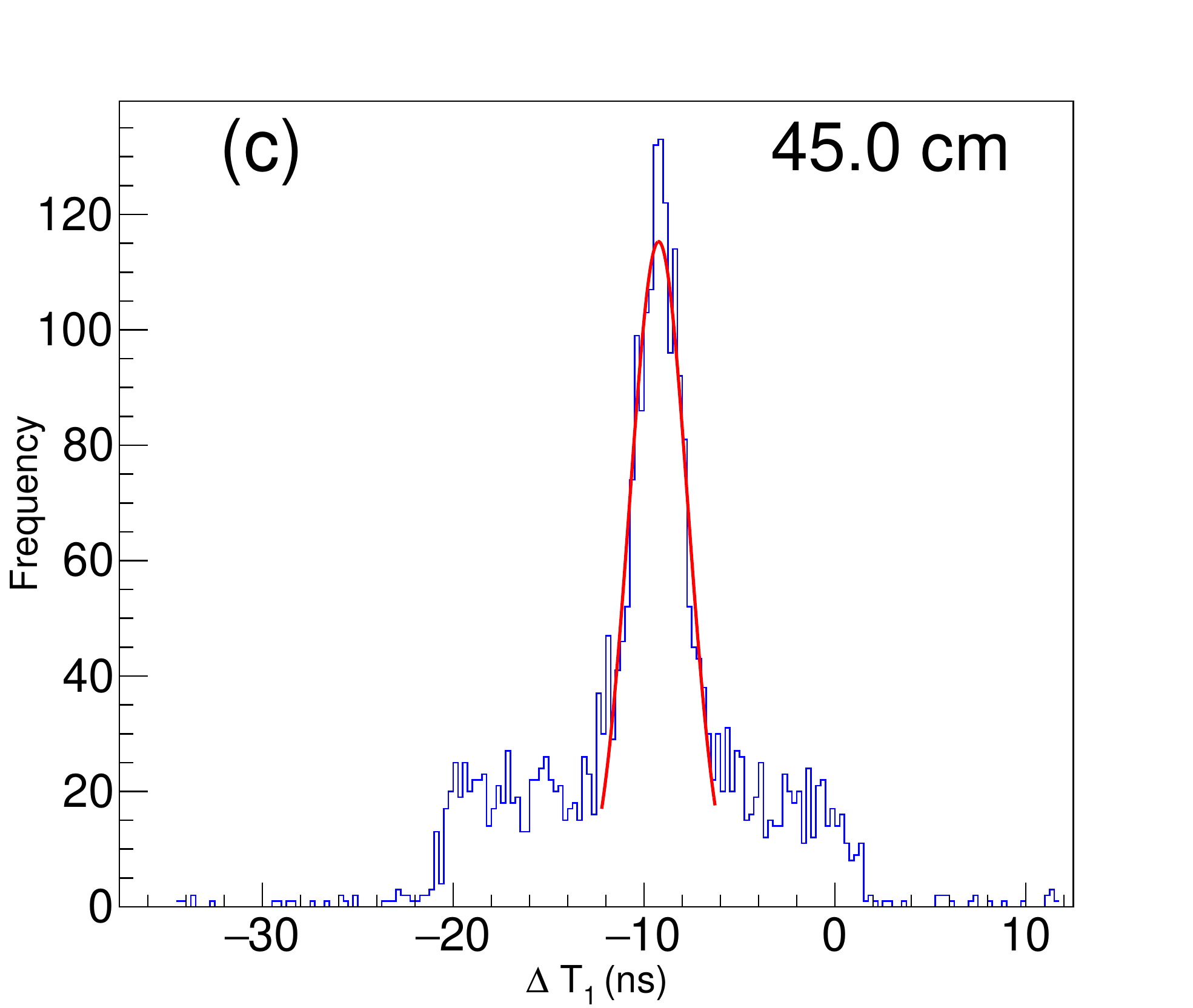} }}
    \subfloat{{\includegraphics[width=7.7cm,height=5.3cm]{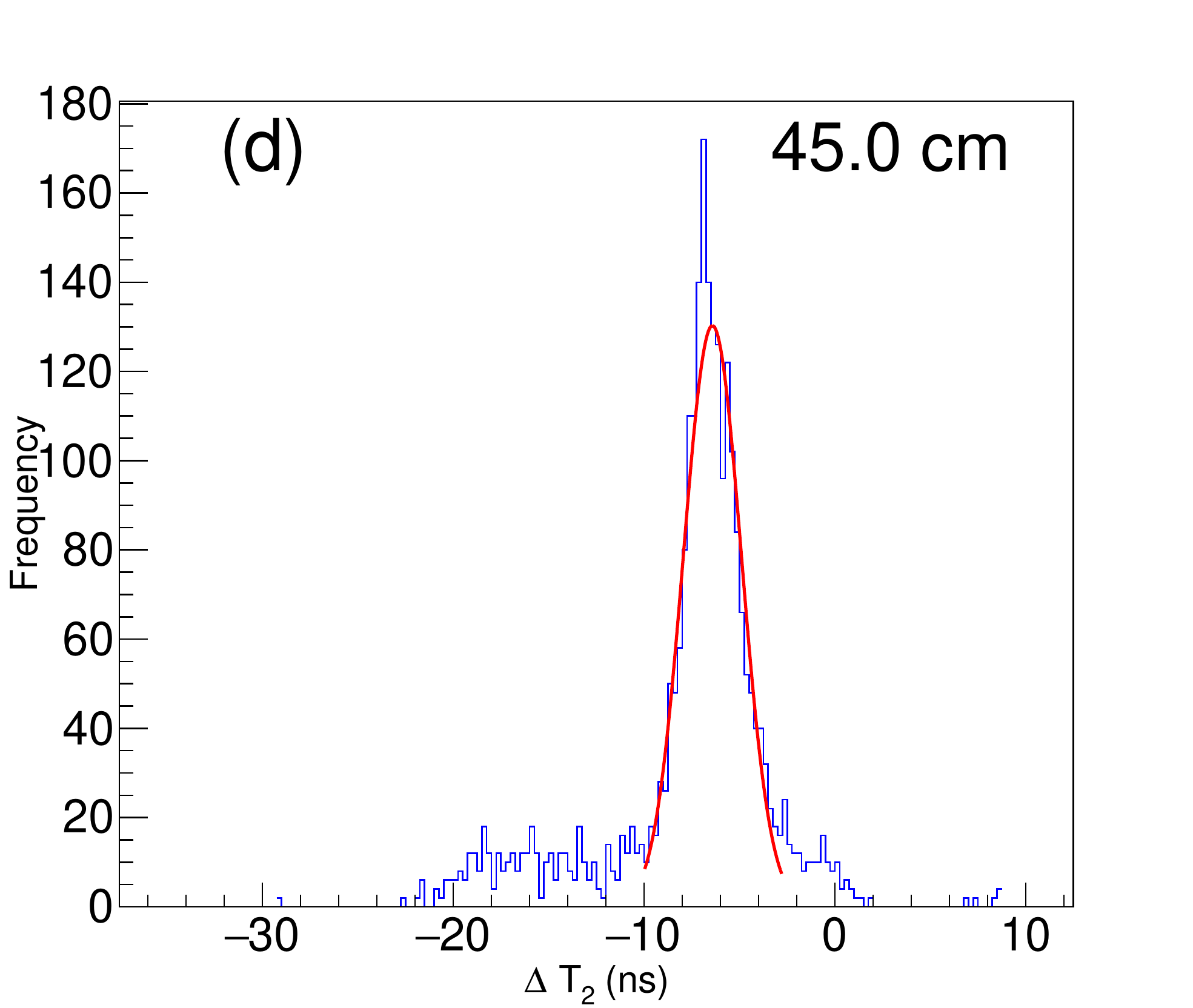} }}\\
    \subfloat{{\includegraphics[width=7.7cm,height=5.3cm]{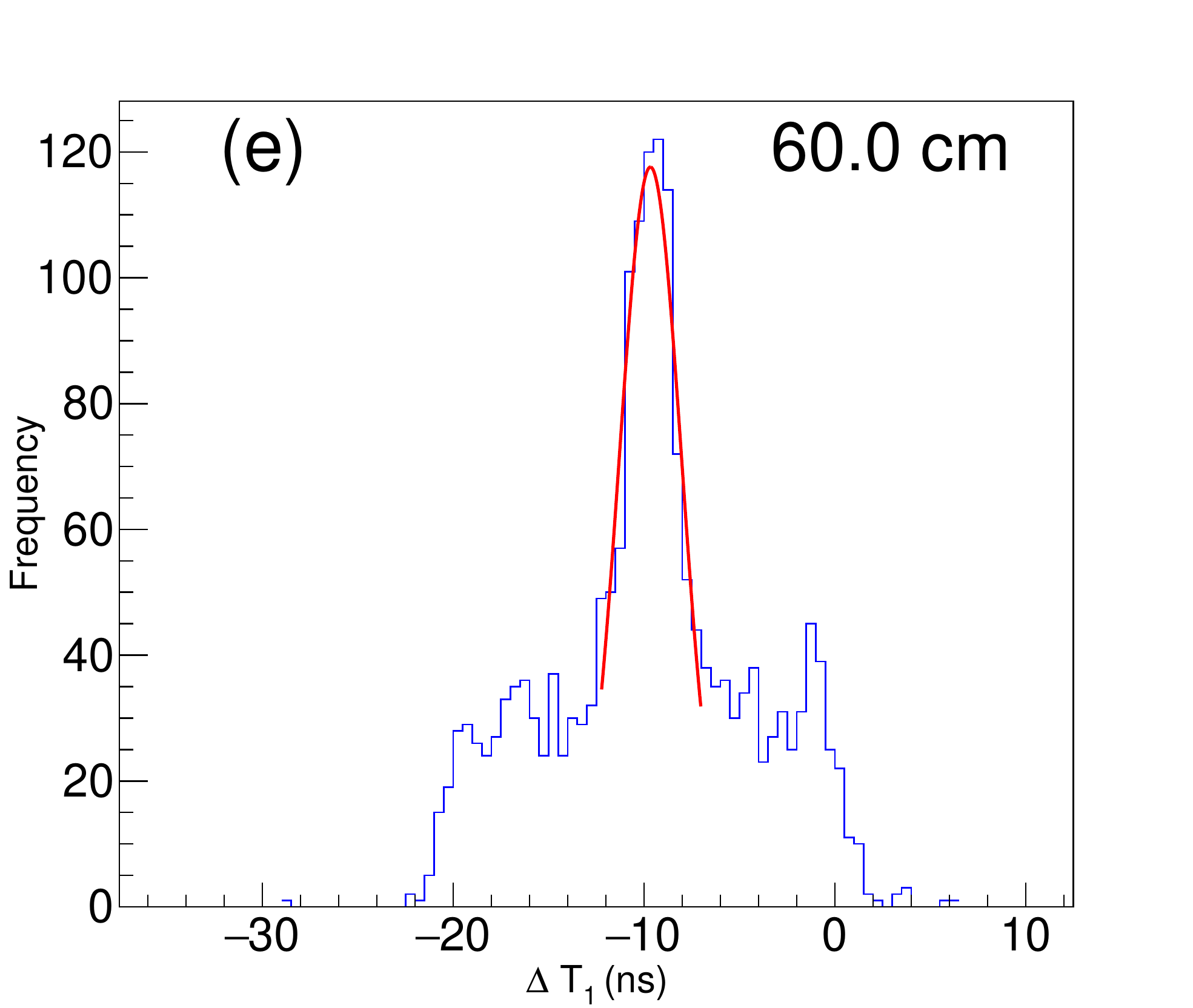} }}
    \subfloat{{\includegraphics[width=7.7cm,height=5.3cm]{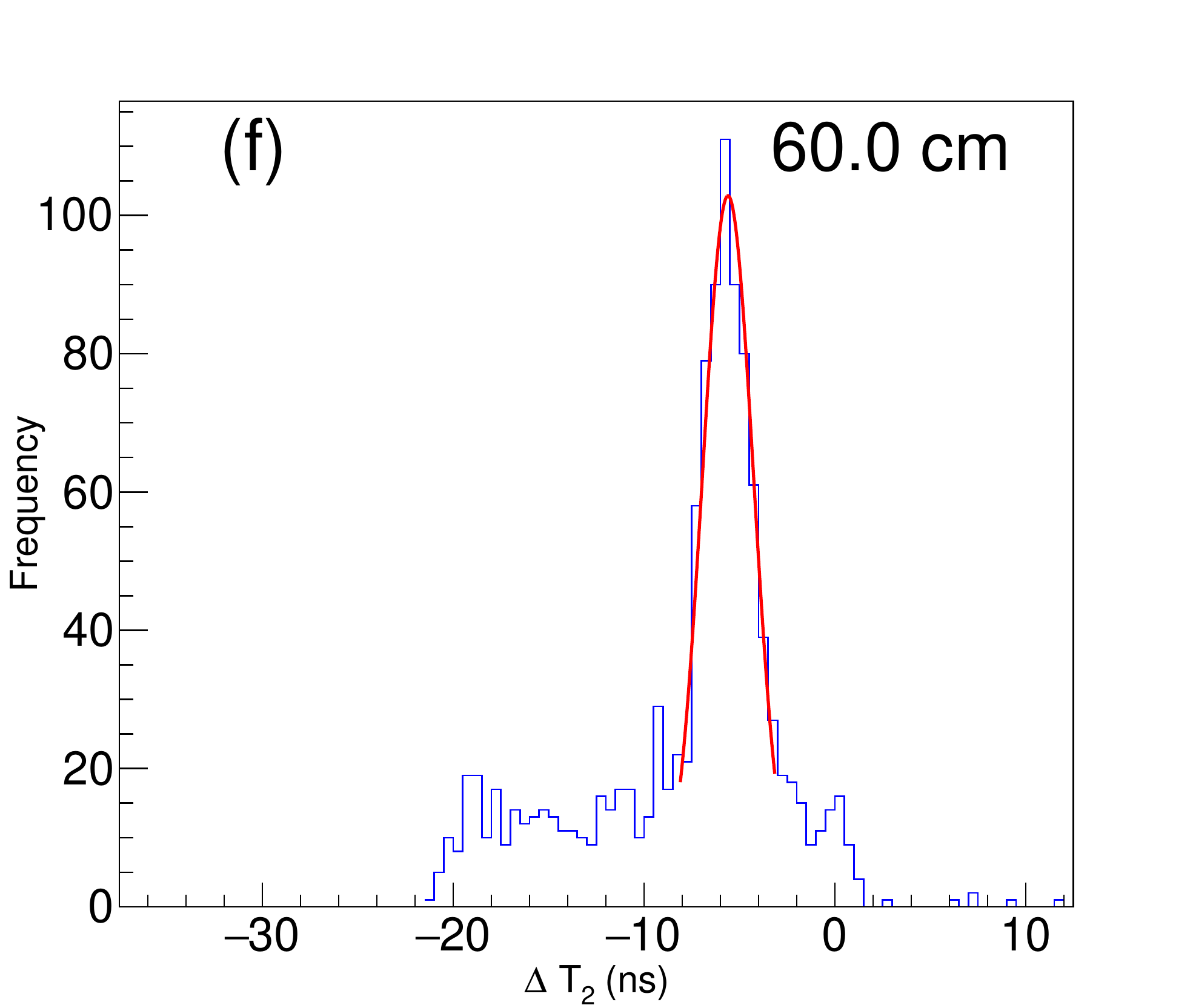} }}\\
    \subfloat{{\includegraphics[width=7.7cm,height=5.3cm]{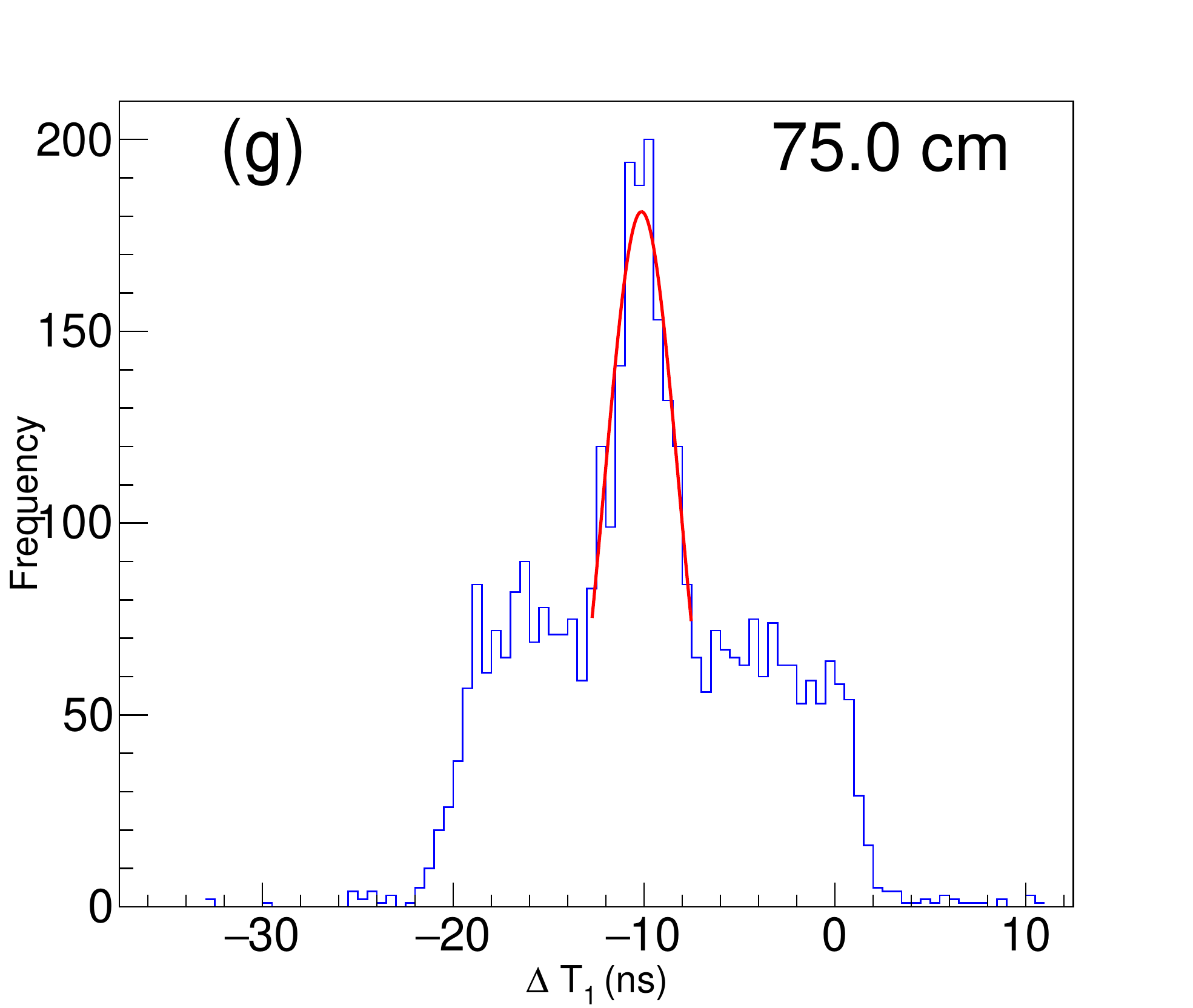} }}
    \subfloat{{\includegraphics[width=7.7cm,height=5.3cm]{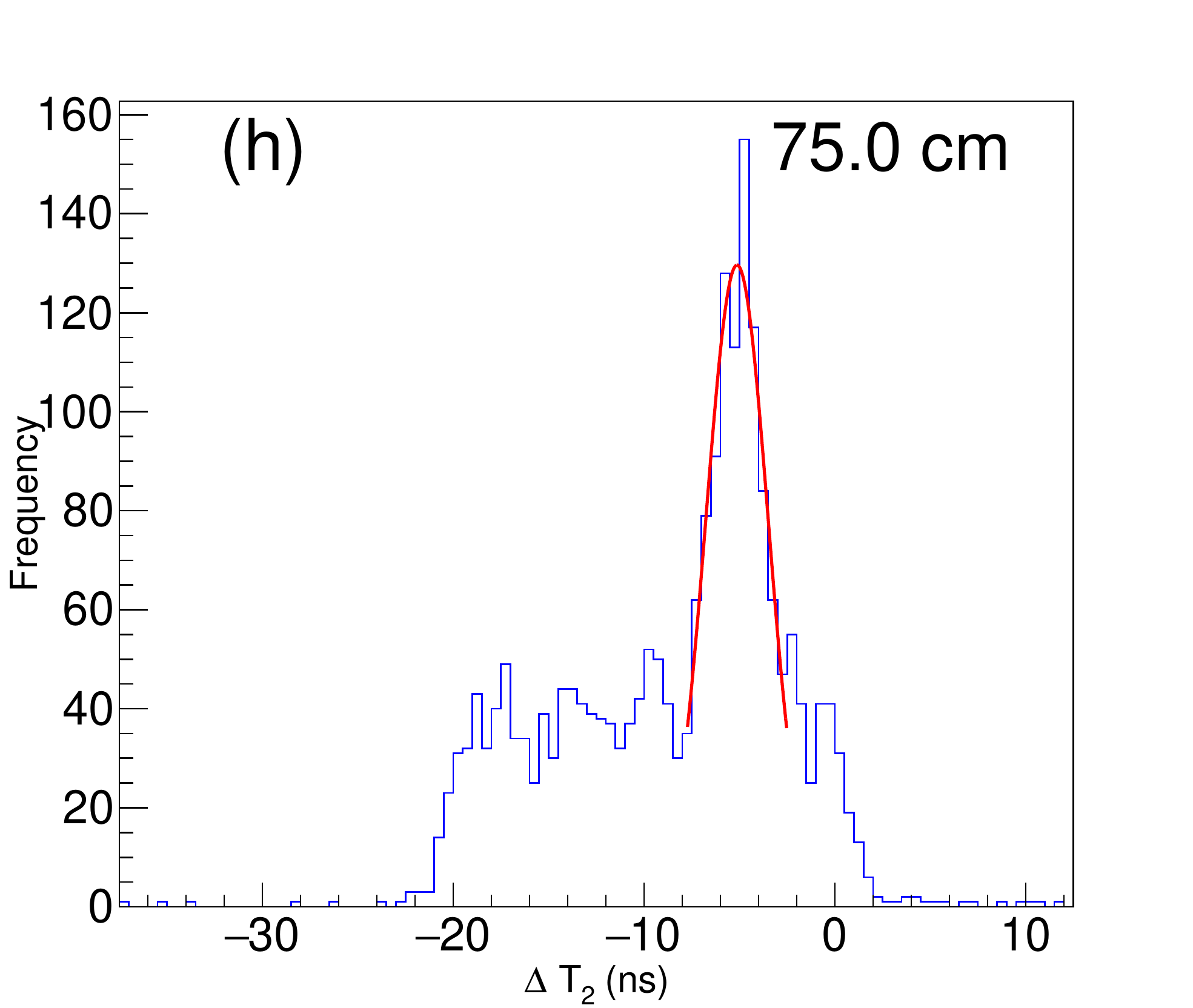} }}
    \caption{$\Delta T$ distributions for various distances between MRPCs.}%
    \label{fig:tdc33}%
\end{figure}

The "X" and "Y" coordinates of hits are recorded along with the time of arrival
of the photon at the detector. We obtain lines of response by joining the hit
coordinates of MRPC1 and MRPC2 only for events with single multiplicity (with
one strip hit per plane per detector). The timing Information infers the exact
position of the source on the line of response. The difference between timings
of two opposite photons is calculated as $\Delta T = t_{MRPC1} - t_{MRPC2}$. There were
offset in cable lengths and the TDC path of different channels. To avoid
this, we have taken two readings for a fixed distance between the MRPCs. First,
the source is kept at the bottom MRPC and we obtain $\Delta T_1 = t_{MRPC1} - t_{MRPC2}$.
The same reading is repeated but with source just below the top MRPC and
we calculate $\Delta T_2 = t_{MRPC1} - t_{MRPC2}$. Finally we calculate the time of 
flight TOF=($|\Delta T_1 - \Delta T_2|$)/2. We took readings for the four values of 
separation (30 cm, 45 cm, 60 cm and 75 cm) between MRPCs. The opearting voltage of MRPCs was set at 15 kV. Our results of time of 
flight calculation for these distances have been summarized in table \ref{table:tof}. 
The figures \ref{fig:tdc33}(a), \ref{fig:tdc33}(c), \ref{fig:tdc33}(e) and \ref{fig:tdc33}(g) show the $\Delta T_{1}$ distributions of the central strips with source at bottom MRPC
 and \ref{fig:tdc33}(b), \ref{fig:tdc33}(d), \ref{fig:tdc33}(f) and
\ref{fig:tdc33}(h)  show the $\Delta T_{2}$ distributions of the central strips with source at top MRPC. 

\begin{figure}[H]%
    \centering
    \subfloat{{\includegraphics[width=15.5cm,height=5.3cm]{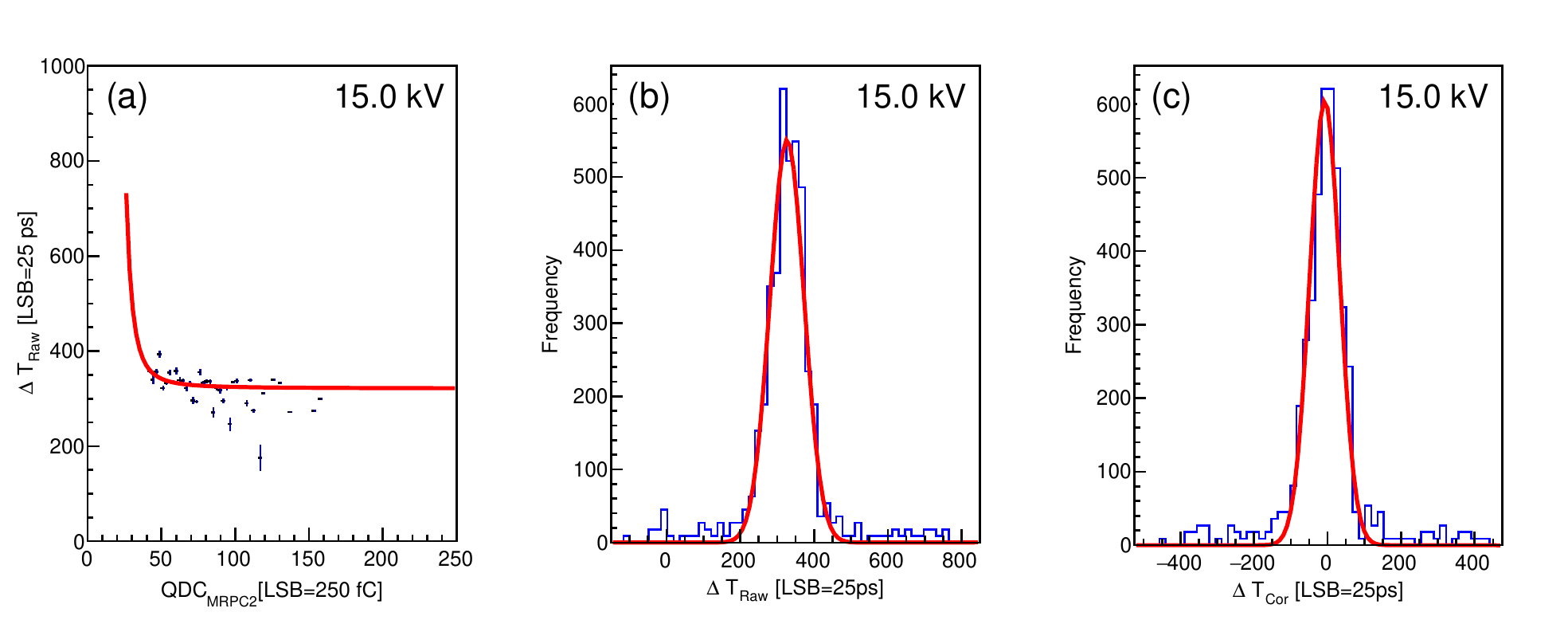} }}\\
    \subfloat{{\includegraphics[width=15.5cm,height=5.3cm]{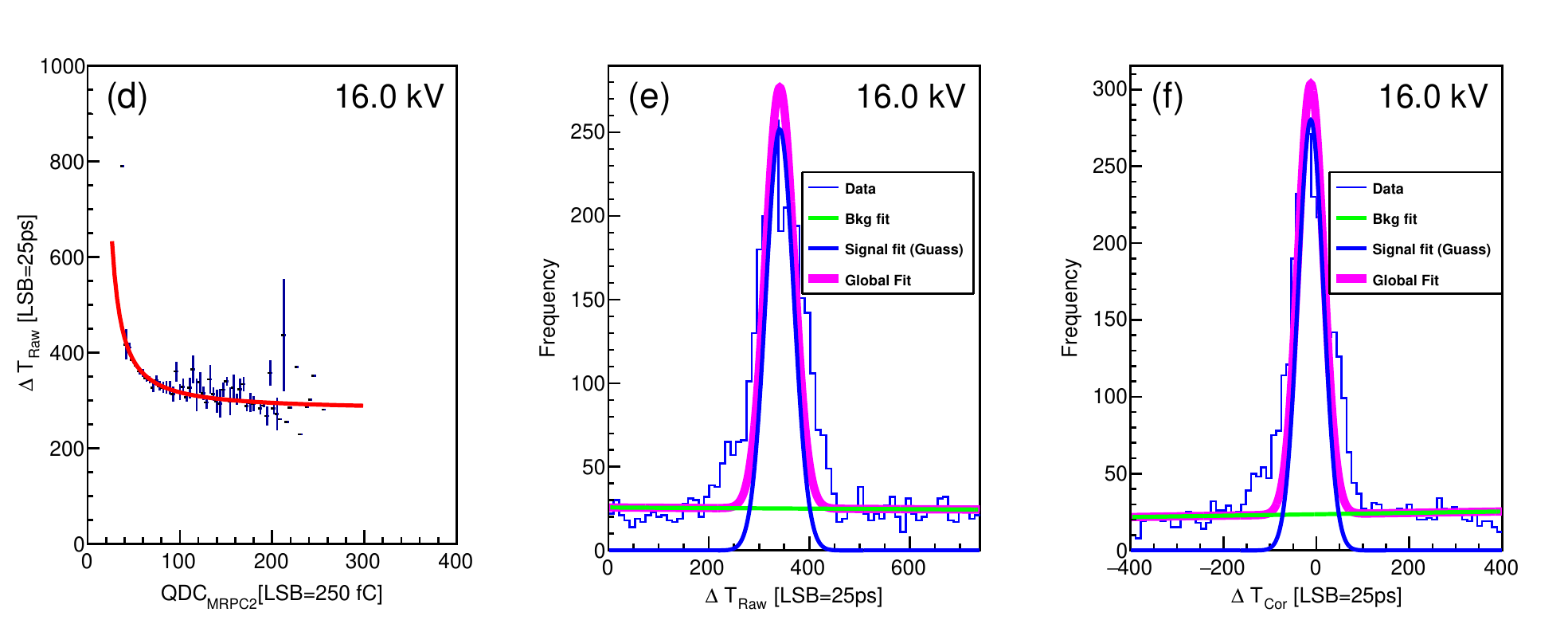} }}\\
    \subfloat{{\includegraphics[width=15.5cm,height=5.3cm]{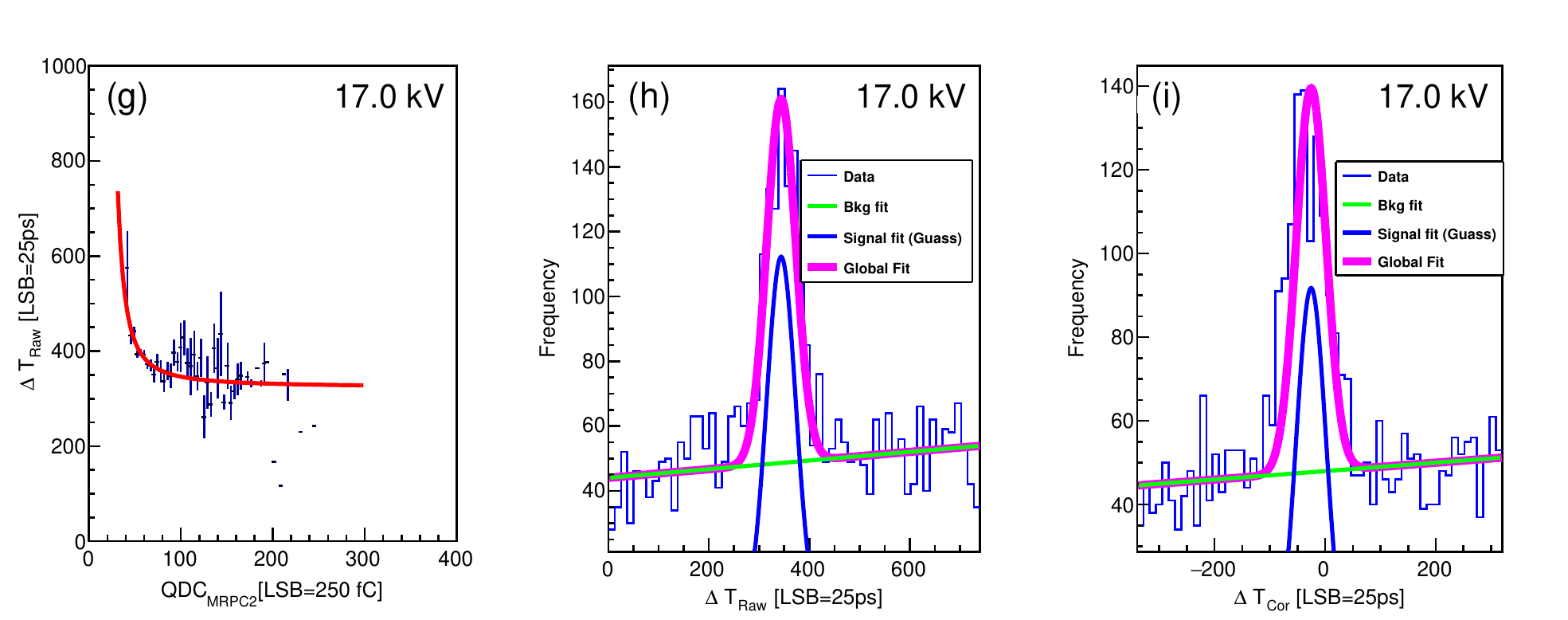} }}\\
    \subfloat{{\includegraphics[width=15.5cm,height=5.3cm]{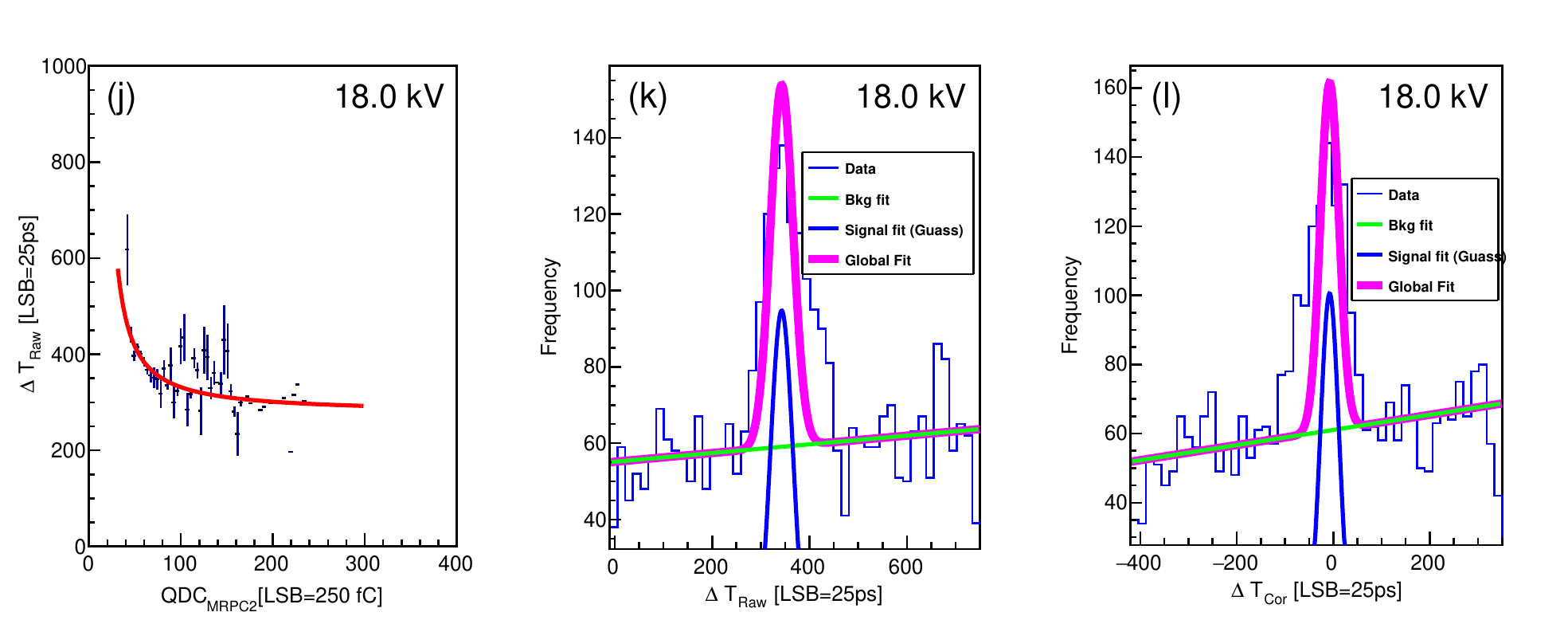} }}
    \caption{First column shows the $\Delta T$ versus $QDC_{MRPC2}$ profile histograms, second column and third column show the raw $\Delta T$ distributions and corrected $\Delta T$ distributions for different operating voltages.}%
    \label{fig:mrpc15}%
\end{figure}

\begin{table}[tbh]
\centering
\begin{tabular}{|c|c|c|c|c|}
\hline
Distance & $\Delta T_1$ (ns) & $\Delta T_2$ (ns) & TOF (ns) & Exp. TOF (ns) \\
\hline
30 cm & -8.38 $\pm$ 0.05 & -6.24 $\pm$ 0.07 & 1.07 $\pm$ 0.06 & 1.0  \\ 
\hline
45 cm & -9.24 $\pm$ 0.04 & -6.42 $\pm$ 0.04 & 1.41 $\pm$ 0.04 & 1.5  \\
\hline
60 cm & -9.67 $\pm$ 0.07 & -5.60 $\pm$ 0.06 & 2.03 $\pm$ 0.07 & 2.0  \\
\hline
75 cm & -10.13 $\pm$ 0.07 & -5.13 $\pm$ 0.07 & 2.50 $\pm$ 0.07 & 2.5  \\    
\hline
\end{tabular}
\caption{Measurement of the time of flight (TOF) for various distances between MRPCs}
\label{table:tof}
\end{table}
\subsection{Time resolution for 511 keV gammas}
The time resolution is the crucial parameter for a TOF-PET device. The Anusparsh boards were used to get 
both the digital as well as analog information of each event. The analog information is required to apply time walk correction 
to the TDC data. Data was taken for four values of operating high voltage 15kV, 16 kV, 17 kV and 18 kV with source placed at the bottom MRPC. The separation between the two MRPCs was kept fixed at $\sim$ 30 cm.
Figures \ref{fig:mrpc15} show the results for the data taken at these four high voltages. The first column shows the profile histograms
of $\Delta T_{Raw}$ versus $QDC_{MRPC2}$. These profile histograms are fitted to a function exp$[-p_0 /x + p_1] + p_2$. The TDC value of each event is then corrected 
by using the fit parameters $p_0$, $p_1$ and $p_2$ and the QDC value. The calibration is done using the equation \ref{eqn1}. The
second column shows the raw $\Delta T$ distributions. The third column shows the corrected $\Delta$T distributions. 
The plots in second and third columns are fitted with a combination of a gaussian and a first order polynomial function. 
The gaussian peak corresponds to the source signal and the linear function fits the background. 
The time resolution values are listed in table \ref{table:tof1}. 
\begin{table}[tbh]
\centering
\begin{tabular}{|c|c|c|}
\hline
H.V. & $\sigma_{T}$ (ps) & $\sigma_{T}$ (ps)  \\
     & (Raw) &  (Corrected) \\
\hline
15 kV & 1157.25 $\pm$ 15.00 & 1022.00 $\pm$ 14.25 \\ 
\hline
16 kV & 757.50 $\pm$ 8.50  & 679.25 $\pm$ 9.00  \\ 
\hline
17 kV & 722.75 $\pm$ 5.00  & 674.50 $\pm$ 7.75  \\
\hline
18 kV & 583.75 $\pm$ 7.50  & 480.75 $\pm$ 5.00 \\ 
\hline
\end{tabular}
\caption[Time resolution at 16 kV and 17 kV]{Time resolution at different high voltages}
\label{table:tof1}
\end{table}
\section{Geant4 simulation of Efficiency for 511 keV gammas}
\begin{figure}[H]%
    \centering
    \subfloat{{\includegraphics[width=7.7cm,height=7.5cm]{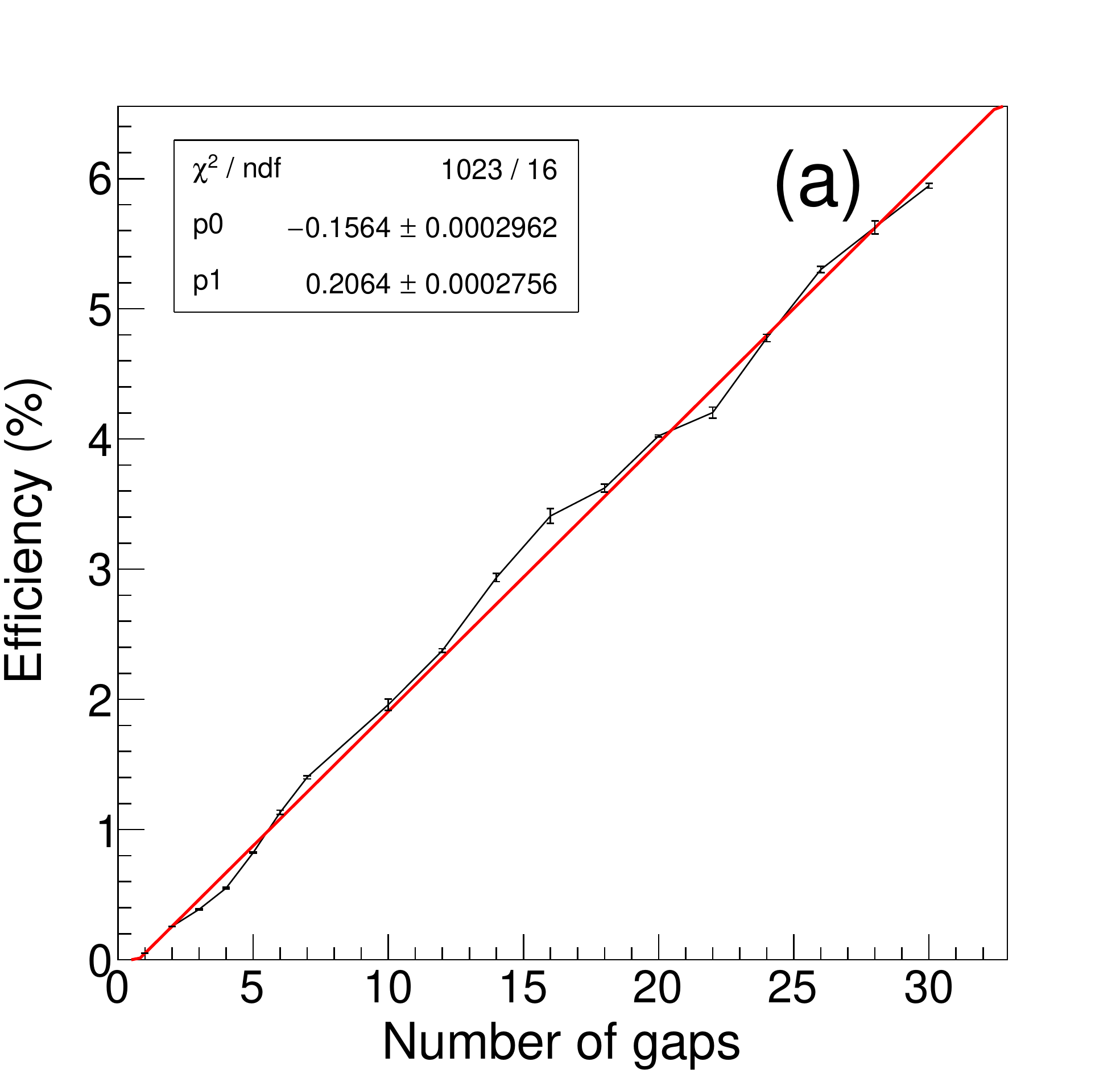} }}%
   % \qquad
    \subfloat{{\includegraphics[width=7.7cm,height=7.5cm]{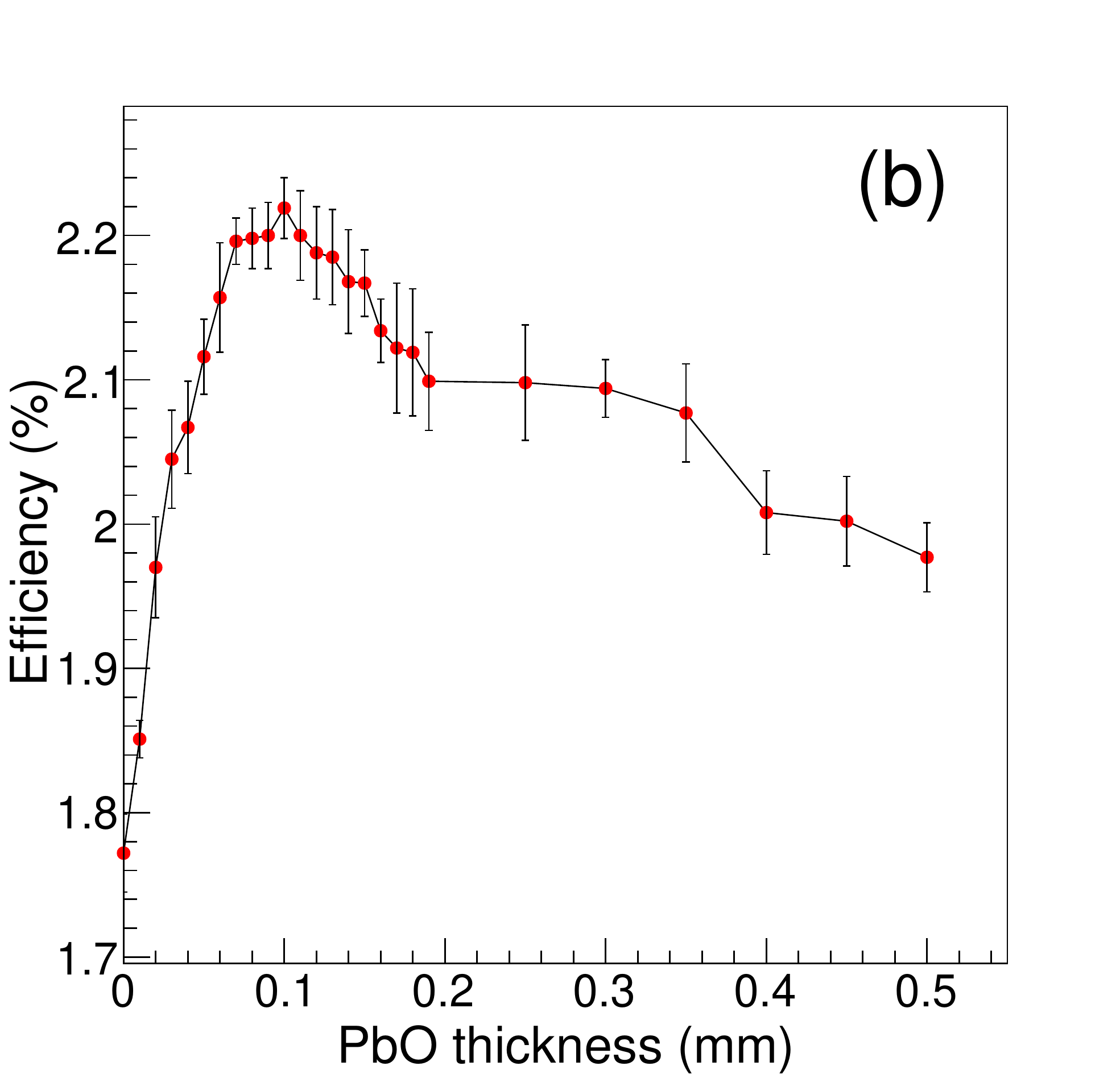}}}% 
    \caption[Efficiency vs thickness of the PbO coat.]{(a) Efficiency vs number of gaps and (b) efficiency vs thickness of the PbO coat.}
\label{fig:mrpc18}
\end{figure}

MRPCs are gas filled detectors and have excellent effeciency for minimum ionising 
particles (muons). Efficiency study of our detector for cosmic muons can be found in \cite{mrpc4}. 
MRPCs have very low efficiency for gammas. Since our experiment involves 511 keV gammas 
produced by ${}^{22}Na$ source, we did a Geant4 simulation study to estimate the efficiency 
of our detector for 511 keV gammas. The MRPC was simulated according to the design details described 
in section \ref{fab}. Gammas of energy 511 keV generated by Geant4 monte carlo were showered 
within a solid angle which covered the entire active detector area. 50,000 photons were used 
to generate a set of data. Ten such data samples were created and the efficiency was calculted for each data set and then their mean and 
standard deviation was calulated. We repeated this exercise by varying the number of gas gaps from 1 to 30.
Figure \ref{fig:mrpc18}(a) shows the plot of effeciency versus the number of gaps in 
the MRPC. The data points are the mean values of efficiency and the errors are given by $\sigma /\sqrt{\rm (N-1)}$ (where "$\sigma$" is the standard deviation and "N" is the number data samples). As it can be seen that the efficiency is very low for 511 keV gammas. 
Our detector has six gaps which corresponds to an efficiency of $\sim$ 1\%. 
Efficency increases with the increase in the number of gaps. 
The efficiency can be improved if a high "Z" material is used as a converter, 
which converts gammas into electrons through photoelectric effect or compton scattering. 
We did a simulation study to estimate the improvement in efficiency by using PbO as a converter material.
A PbO coat was applied on the inner side of an outside electrode. 
The thickness of the coat was varied from 0.00 mm to 0.50 mm. 
Ten data sets were obtained for each thickness by shooting 50,000 photons on the coated electrode side of the MRPC. 
Efficiency was obtained in similar way as obtained above. Figure \ref{fig:mrpc18}(b) shows 
the efficiency versus PbO coat thickness plot. The data points are the mean values obtained 
from the ten data sets and the errors are given by $\sigma /\sqrt{\rm (N-1)}$. We see an improvement 
in effeciency near 0.1 mm thickness of the coat. The number of gaps (6 gaps) was kept fixed for this study.
\section{Conclusion} 
We developed and characterized several six-gap glass MRPCs and extensively studied their performance over a long period of time.
We  measured Time Of Flight (TOF) of 511 keV gammas produced by (${}^{22}Na$) source between the two MRPCs for different
distances between the two detectors. The measured TOF is in good agreement with the expected values. 
We studied the 
time resolution of our detector for 511 keV photons at different operating high volatges. The time resolution improves with incresing high voltage. We could go upto 18 kV and the time resolution at this high voltage 
is $\sim$ 480 ps which includes electronic jitter of $\sim$ 120 ps. 
The time walk correction using the analog information from the Anusparsh boards significantly improves the time
resolution. 
We did Geant4 simulation to estimate the efficiency of our detector for 511 keV gammas for different number of gaps in an MRPC. The efficency increases
with increasing number of gas gaps. 
The efficiency of our detector can also be improved by using a PbO coat on the inner side
of one of the outside electrodes.

\acknowledgments

We would like to thank Dr. Moon Moon Devi who provided valuable help in the initial stages
of this work. We also thank Prof. V. M. Datar, Prof. Amol Dighe and
Dr. B.K. Nayak for their valuable feedbacks. Thanks are also due to Dr. V. B. Chandratre, Ms. Meneka Sukhwani and Mr. S. Joshi for their help in designing, fabricating and
 intergrating the Anusparsh boards into our experimental setup.

\end{document}